\documentclass[11pt]{article}

\usepackage[]{graphicx,float,latexsym,times}
\usepackage{amsfonts,amstext,amsmath,amssymb,amsthm}
\usepackage{caption2}

\RequirePackage[OT1]{fontenc}
\long\def\symbolfootnote[#1]#2{\begingroup%
	\def\thefootnote{\fnsymbol{footnote}}\footnote[#1]{#2}\endgroup}

\usepackage{titlesec} 

\titleformat{\section}{\large\bfseries}{\thesection.}{.5em}{}
\titlespacing*{\section}{0pt}{*3}{*2}
\titleformat{\subsection}{\normalfont\bfseries}{\thesubsection.}{.5em}{}
\titlespacing*{\subsection} {0pt}{*3}{*2}
\titleformat{\subsubsection}{\normalfont\bfseries}{\thesubsubsection.}{.5em}{}
\titlespacing*{\subsubsection} {0pt}{*3}{*2}

\usepackage{xspace}
\usepackage{geometry}
\usepackage{array}
\usepackage{subfigure}
\usepackage{epsfig}
\usepackage{hhline}
\usepackage{bbm}
\usepackage{multirow}
\usepackage{dcolumn}
\usepackage{color}
\usepackage{lscape}
\usepackage{hyperref}

\usepackage{colortbl, xcolor}
\definecolor{Gray}{gray}{0.9}

\theoremstyle{plain} 
\newtheorem{theorem}{Theorem}[section]
\newtheorem{lemma}{Lemma}[section]
\newtheorem{corollary}{Corollary}[section]

\newtheorem{proposition}{Proposition}[section]

\theoremstyle{definition} 

\numberwithin{equation}{section} 

\usepackage[authoryear]{natbib}

\DeclareMathAccent{\widehat}{\mathord}{largesymbols}{"62}
\DeclareMathAccent{\widetilde}{\mathord}{largesymbols}{"65}

\def\eb{\textrm{\mathversion{bold}$\mathbf{\beta}$\mathversion{normal}}}

\def\eO{\textrm{\mathversion{bold}$\mathbf{\Omega}$\mathversion{normal}}}

\def\eE{\mathbb{E}}

\def\e1{1\!\!1}

\def\rr{\textrm{\mathversion{bold}$\mathbf{r}$\mathversion{normal}}}

\numberwithin{equation}{section} 

\def\ee1{\textrm{\mathversion{bold}$\mathbf{\varepsilon}$\mathversion{normal}}}

\def\eu{\mathbf{{u}}}
\def\eR{\mathbf{{R}}}

\newcommand{\N}{\mathbb{N}}
\newcommand{\R}{\mathbb{R}}
\newcommand{\PP}{\mathbb{P}}

\def\eX{\mathbf{X}}
\def\eJ{\mathbf{J}}

\newcommand{\Var}{\mathbb{V}\mbox{ar}\,}

\def\argmin{\mathop{\mathrm{arg\,min}}} 
\def\hh{ \hspace*{0.5cm}}
\begin{document}
\title {{\bf{\Large Real-time detection of a change-point in a linear expectile model}}}
\date{}
\maketitle
 \author{
 	\begin{center}
 		\vskip -1cm 
 		Gabriela CIUPERCA \\
\small   {Institut Camille Jordan,  UMR 5208, Universit\'e Claude Bernard Lyon 1, France}
\end{center}
}
\symbolfootnote[0]{\normalsize Address correspondence to  G. Ciuperca,
	Universit\'e de Lyon, Universit\'e Claude Bernard Lyon 1, CNRS, UMR 5208, Institut Camille Jordan, Bat.  Braconnier, 43, blvd du 11 novembre 1918, F - 69622 Villeurbanne Cedex, France; \textit{E-mail}: Gabriela.Ciuperca@univ-lyon1.fr}
{\small \noindent\textbf{Abstract:} 
	In the present paper we address the real-time detection  problem of a change-point in the coefficients of a linear model with the possibility that the model errors are asymmetrical and that the explanatory variables number is large. We build  test statistics   based on the cumulative sum (CUSUM)  of the expectile function derivatives calculated on the residuals obtained by the expectile and adaptive LASSO expectile estimation methods. The asymptotic distribution of these statistics are obtained under the hypothesis that the model does not change. Moreover, we prove that they diverge when the model changes at an unknown observation.  The asymptotic study of the  test statistics  under these two hypotheses allows us to find the asymptotic critical region and the stopping time, that is the observation where the model will change.  The empirical performance is investigated by a comparative simulation study with other statistics of   CUSUM type. Two examples on real data are also presented to demonstrate its interest in practice.
	 \\
}
{\small \noindent\textbf{Keywords:} real-time change-point detection; asymmetric model error; expectile; adaptive LASSO; stopping time.}\\
{\small \noindent\textbf{Subject Classifications:} 62F05; 62J07; 62L10}

 
\section{Introduction} 
\label{Intro} 
In many applications (medicine, finance, engineering, ecology, meteorology) it is necessary to solve the problem of testing in real-time of a change in a model which does not satisfy classical conditions. Moreover, with the recent technological advancements that allow the collection of a vast amounts of data, the studied model can contain a large number of explanatory variables. These involve the need to  automatically select  the relevant variables simultaneously with the change detection, whence the practical interest of the results obtained in this paper.\\
Therefore, we consider the real-time detection of a change-point in a linear model when the model errors can be asymmetrical and when the number of explanatory variables can be large.\\
There are two types of change-point detection: a posteriori and in real-time (sequential). The a posteriori detection is done once the experiment has ended, after which we ask whether there have been changes in the model and if so, find their number and their location. Real-time detection is done with each observation: we have a model based on historical data and we know that the change did not occur. At each new observation we test whether the model has changed or not. The test is based on a statistic built on the residuals corresponding to the estimators obtained on the  historical data model. If the model errors satisfy the classical conditions, that is zero mean and bounded variance, \cite{Horvath.Huskova.Kokoszka.Steinebach.04} introduced the test statistic based on the cumulative sum (CUSUM) of the residuals obtained by the least squared (LS) estimation method. However, if the law of errors is asymmetrical, then the LS estimation method is not appropriate to estimate the model parameters because it produces  non accurate estimators.  Then, \cite{Newey-Powell.87} introduced the expectile estimation method which is more appropriate when the first four moments of the model errors $\varepsilon$  exist and the distribution of   $\varepsilon$ is asymmetrical. Another possible estimation method when $ \varepsilon $ does not satisfy the classical conditions on the moments is the quantile one (see \cite{Koenker.05}). The expectile method has the advantage over the quantile method that it is differentiable, therefore theoretical studies  and numerical computation are easier. Another advantage is that the asymptotic variance of the expectile estimator can be calculated without going via the density of the model error. \\
If the number of explanatory variables of the model is large and only a part is relevant (with non-zero coefficients) then the automatic selection of these variables can be carried out using the adaptive LASSO penalty introduced by \cite{Zou.06} for the LS loss function. If the loss function is  quantile or expectile, then, corresponding adaptive LASSO estimators have been proposed and studied by \cite{Ciuperca-16} and \cite{Liao-Park-Choi.19}, respectively. \\
Let us give some recent paper  references  where the CUSUM method or variations have been used for detection in linear models of a change-point in real-time.  \cite{Zou-Wang-Tang.15} propose and study a test statistic as a  CUSUM of the subgradient of the quantile process in order to detect in real-time a change in a linear quantile model. \cite{Kirch-Weber.18} propose different statistics with respect to the number of observations included in the partial sum: modified moving-sum-statistic, Page’s cumulative-sum-statistic  and the standard moving-sum-statistic.
 \cite{Zhang-Li.17} consider a model for which  the effect of a covariate on the response
 variable is linear but varies below and above an unknown threshold in a continuous way. Therefore, \cite{Zhang-Li.17} study a different model from that of the present paper because they consider that the change occurs when one of the regressors exceeds a certain value which is seen as the change sought and estimated. Based on a weighted CUSUM type statistic, they develop a testing procedure for the existence of structural change at a given expectile. The same model where the change occurs with respect to an unknown value of a covariable is considered by \cite{Zhang-Li.17b} which  develop a score-like test  based on a weighted CUSUM process. \cite{Jiang.Kurozumi.20} applied the CUSUM test based on LS residuals to sequentially detect structural change in a linear model  with a trend. For the proposed test statistic, they get the limiting null distribution and the divergence under the alternative. In addition,   the asymptotic distributions of the corresponding stopping times is derived. On the other hand, \cite{Jiang.Kurozumi.19} investigated two modified versions of the CUSUM test    to avoid losing power when the mean of the regressor is orthogonal to the shift. In  \cite{Qi-Duan-Tian-Li.16}, a new class of fluctuation sequential tests based on recursive estimates is  proposed, for which asymptotic behaviour is studied. For the  change-point detection on the parameter of a certain  discrete-time  stochastic processes, \cite{Nedenyi-18} presents a statistic based on the CUSUM of the estimates of certain martingale difference sequences. 
 In order to  locate the change-point in a  multivariate data with heavy-tailed distribution, \cite{Liu-Zhou-Zhang.19} propose a  tail adaptive approach. Moreover,  in order to detect  in real-time  an abrupt change in linear regression models, \cite{Geng-Zhang-Huie-Lai.19} propose a novel algorithm, in Bayesian and non-Bayesian formulation. Even if it is not the CUSUM method that is used, we consider it important to cite \cite{Horvath-Rice.19}'s paper where a linear factor model is considered and  where the dimension of the factors and the sample size tend jointly to
 infinity. For testing the structural stability of the  model, it is proved that, under the null hypothesis,  if the effect of the factors is sufficiently strong then the processes of partial
 sample estimates of the largest eigenvalues of the sample covariance matrix   have Gaussian limits. Works which simultaneously consider the   relevant variables detection  and also the   real-time detection of a change in the model by a CUSUM type technique are \cite{Ciuperca-15} and \cite{Ciuperca-18} where the  loss  function is LS and quantile, respectively, with an adaptive LASSO type penalty.\\
 To the best of the author's knowledge, there is  no work in the literature regarding the real-time detection of a change in the coefficients of a linear model using the expectile method and when the coefficient number is large. In this last  case, the adaptive LASSO expectile method   will be used to select   the relevant variables  on the historical data.  \\
 Finally, but not lastly, we would like to remind that the CUSUM-type methods are used to detect a change in the parameters of a time series. Here are some references from the  recent literature: \cite{Song-Kang.20}, \cite{Gronneberg-Holcblat.19},  \cite{Praskova-18}, \cite{Chen-Hu.17}. \\
 The remainder of the paper is organized as follows. In Section \ref{sect_models} we introduce the model, the assumptions, the null and alternative hypothesis for the model and we state  two auxiliary results. In order to  detect  the change in the model in real-time, two test statistics are proposed and asymptotically studied under the null hypothesis (no change)  and under the alternative hypothesis (there exist a change in the model coefficients after the historical data) in Section \ref{sect_results}. In Section \ref{simus}, we study and compare by a numerical study the proposed statistics with other test statistics. The two test statistics are also applied on two real data. The proofs of the results are postponed to  Section  \ref{sect_proof}. 
 	
\section{Preliminaries and models}
\label{sect_models}
In this section we present the model, the considered assumptions and some preliminary theoretical results.\\

First of all,  give some general notations  that will be used throughout the paper. We make the convention that all vectors and matrices are in bold and that all vectors are column. For a vector $\textbf{v}$, we denote by $\textbf{v}^\top$ its transposed, by $\|\textbf{v}\|_1$,  $\|\textbf{v}\|_2$ and $\|\textbf{v}\|_\infty$, the $L_1$, $L_2$, $L_\infty$ norms, respectively. For $q \in \N$, the $q$-vector $\textbf{0}_q$ has all components $0$.  We denote by $C$ a positive  generic constant  independent from $m$, which value may   differ from one formula to another. The value of $C$ is not of interest. For a set ${\cal A}$, let us denote by $|{\cal A}|$  the cardinality of ${\cal A}$ and by  ${\cal A}^c$ its complementary set.  The notations $ \overset{\cal L} {\underset{m \rightarrow \infty}{\longrightarrow}}$, $ \overset{\PP} {\underset{m \rightarrow \infty}{\longrightarrow}}$ represent respectively the convergence in distribution and in probability, as $m \rightarrow \infty$.  \\ 

We  consider a model on $ m $ observations:
\begin{equation}
\label{eq1}
Y_i=\eX_i^\top \eb+\varepsilon_i, \qquad i=1, \cdots , m. 
\end{equation}
The observations $1, \cdots , m$ are called historical observations and $(Y_i,\eX_i)_{1 \leqslant i \leqslant m}$ historical data. 
After these observations, other $T_m$ incoming observations sequentially are measured for the response variable $Y$ and for the vector $\eX$ which contains the explanatory variables. The considered model on these observations is:
\begin{equation}
\label{eq10}
Y_i=\eX_i^\top \eb_i+\varepsilon_i, \qquad i=m+1, \cdots , m+T_m. 
\end{equation}
For models (\ref{eq1}) and (\ref{eq10}), the $p$-vector of the explanatory variables $\eX_i$ is deterministic, with the components $X_{ij}$, for $j=1, \cdots , p$ and $i=1, \cdots , m+T_m$.  Furthermore, the values of $(Y_i, \eX_i)$ are known for any $i=1, \cdots, m, m+1, \cdots ,m+T_m$.\\
For model (\ref{eq1}), the parameter vector   $\eb=(\beta_{1}, \cdots , \beta_{p})^\top \in \R^p$ is of dimension $p$ fixed and its true value (unknown) is  $\eb^0$ which does not depend on $m$. Regarding the $p$-vector of parameters $\eb_i$ of model (\ref{eq10}), details will be given later.\\

Through the paper, the following assumptions are considered for the design $(\eX_i)_{1 \leqslant i \leqslant m+T_m}$ :
\begin{description}
\item \textbf{(A1)} $\max_{1 \leqslant i \leqslant m+T_m} \| \eX_i\|_2 \leq C_0$, for some constant $C_0 >0$.
\item \textbf{(A2)} For  $n \in \N$, $  n \leq m+T_m$, and the $p$-square matrix $\eO_n \equiv n^{-1} \sum^n_{i=1} \eX_i \eX_i^\top $,  there exists a positive definite matrix $\eO$, such that, $\lim_{ n\rightarrow \infty} \eO_n =\eO $.
\end{description}  
 
We emphasize that Assumption (A1) was also considered   for  expectile models for example  by \cite{Zhao-Chen-Zhang.18} and \cite{Ciuperca.19}, while Assumption (A2) is standard for linear models to ensure the identifiability of the coefficients (see for example \cite{Zou.06}, \cite{Geng-Zhang-Huie-Lai.19}, \cite{Liao-Park-Choi.19}).\\
Moreover, the errors $(\varepsilon_i)_{1 \leqslant i \leqslant m+T_m}$ of models (\ref{eq1}) and (\ref{eq10}) will be assumed to be of the same distribution, not necessarily symmetrical. In order to introduce the suppositions on  $\varepsilon$, with $\varepsilon$ a generic term of  $\varepsilon_i$, and to perform  the statistical inference proposed in this paper, we introduce the expectile function. For a fixed expectile index  $\tau \in (0,1)$,  the expectile function of order $\tau$ is defined by:
\[\rho_\tau(x)=|\tau - \e1_{x <0}|x^2, \qquad \textrm{for} \quad x \in \R.
\] 
The derivative of $\rho_\tau(x-t)$ in $t=0$ is $g_\tau(x) \equiv \rho'_\tau(x-t)_{t=0}= 2 \tau x \e1_{x \geq 0}+2(1-\tau)x \e1_{x<0}$ and the second derivative is $h_\tau(x) \equiv \rho''_\tau(x-t)_{t=0}=2 \tau \e1_{x \geq 0}+2(1-\tau) \e1_{x<0}$. \\

Thereby, for the errors   $(\varepsilon_i)_{1 \leqslant i \leqslant m+T_m}$, we suppose the following assumption:
\begin{description}
\item \textbf{(A3)} $(\varepsilon_i)_{1 \leqslant i \leqslant m+T_m}$ are i.i.d. continuous,  such that $\eE[\varepsilon^4]< \infty$ and $
\eE[ g_\tau(\varepsilon)]=0$.
\end{description}  
 
Note that assumption (A3) is standard for the  expectile models (see \cite{Liao-Park-Choi.19}, \cite{Zhao-Chen-Zhang.18}, \cite{Gu-Zou.16}, \cite{Ciuperca.19}).\\

We introduce the following notations: 
\[
\underline{c}=\min(\tau, 1-\tau), \quad \bar{c}= \max (\tau, 1-\tau), \quad \mu_{h }=\eE[ h_\tau(\varepsilon)]>0.
\]
 The mean  $\mu_{h }$ depends on the index $\tau$ but, in order to relieve the notation, we drop    $\tau$. \\

Since the errors $(\varepsilon_i)_{1 \leqslant i \leqslant m+T_m}$ satisfy assumption (A3) and not the classical assumptions   of zero mean and bounded variance, \cite{Newey-Powell.87} introduced the expectile estimator for the coefficients of a linear model.  Thus, the parameter vector $\eb$ of model (\ref{eq1}) is estimated by the expectile estimator which is defined by:
\begin{equation}
\label{hbm}
\widehat{\eb}_m \equiv \argmin_{\eb \in \R^p} \sum^m_{i=1} \rho_\tau(Y_i-\eX_i^\top \eb).
\end{equation}
Under assumptions (A1)-(A3), the expectile estimator $\widehat{\eb}_m$ converges to the true value  $\eb^0=(\beta^0_1, \cdots , \beta^0_p)^\top$ with a  convergence rate of order $m^{-1/2}$ (see \cite{Ciuperca.19}). We denote the components of the vector $\widehat{\eb}_m$ by $(\widehat{\beta}_{m,1}, \cdots ,\widehat{\beta}_{m,p} )^\top$.\\
The following proposition  states a more precise result on the asymptotic behavior of $\widehat{\eb}_m$. 

\begin{proposition}
\label{Propositon 1}
Under assumptions (A1)-(A3), we have:
\[
\widehat{\eb}_m =\eb^0 + \mu_{h }^{-1} \eO^{-1} \frac{1}{m} \sum^{m}_{i=1} g_\tau(\varepsilon_i) \eX_i +o_{\PP}(m^{-1/2}).
\]
\end{proposition}
The proof of Proposition  \ref{Propositon 1} is given in Section \ref{sect_proof}.\\

We are now focusing on model (\ref{eq10}), for which two hypotheses are considered, the observations $m+1, \cdots , m+T_m$ constituting the monitoring period.\\
For this model, we test the null hypothesis that after the historical observations, the coefficients of model (\ref{eq10}) coincide with those of  model (\ref{eq1}):\\

\hh $ H_0: \, \eb_i= \eb^0$, for all $i=m+1, \cdots , m+T_m$,\\

\noindent against the alternative hypothesis, that there is an unknown observation $m+k^0_m$ from which there is a change in the coefficients of the model:\\

 \hh $ H_1$:  there exists $k^0_m  \in \{ m+1, \cdots m+T_m\}$,  such that $\left\{ \begin{array}{lll}
\eb_i =\eb^0, & & i=m+1, \cdots , m+k^0_m, \\
\eb_i=\eb^1, & & i=m+k^0_m+1, \cdots, m+T_m,
\end{array}  \right.$ \\ 

\noindent with $\eb^1 \neq \eb^0$ and $ \eb^1$ unknown. Unlike $\eb^0$, the value of $\eb^1$ can depend on $m$. \\

\noindent For the monitoring period, there are two possible cases depending on the values of the number $T_m$ of observations after the historical data. These cases are referred as the following procedures:
 \begin{itemize}
	\item \textit{open-end procedure}, when  $T_m = \infty$ or when ($T_m < \infty$, $\lim_{m \rightarrow \infty}T_m/m=\infty$);
	\item \textit{closed-end procedure}, when $\lim_{m \rightarrow \infty}T_m/m=T \in (0, \infty)$.
\end{itemize} 
\vspace{0.25cm}
For the open-end procedure the monitoring is carried out to infinity if no change occurs in the model, while for the closed-end procedure, the monitoring is stopped after a fixed number of observations even if no change occurs.\\

Under hypothesis $H_1$, the unknown change-point $k^0_m$ can depend on $m$. If $k^0_m$ depends on $m$, thus, it is not very far from the last observation $m$ of the historical data. thus, we consider that $k^0_m$ satisfies the following assumption:
 \begin{description}
 	\item \textbf{(A4)}  $k^0_m=O(m^s)$, with the constant $s$ such that $s>1$ for the open-end procedure and $0 \leq s \leq 1$ for the closed-end procedure. 
 \end{description}  
 This is a typical condition on $k^0_m$   in a real-time detection problem of a change-point (see \cite{Huskova.Kirch.12}, \cite{Ciuperca-18}, \cite{Jiang.Kurozumi.20}). \\
Moreover, assumption (A4) is in accordance with the definition of the two procedures.\\
 
Taking into account the  convergence rate of $\widehat{\eb}_m$ towards $\eb^0$, we consider the following random    processes:
\begin{align*}
\eR_i(\eu)& \equiv \bigg(g_\tau \big(Y_i -\eX_i^\top(\eb^0 +m^{-1/2} \eu)\big)- g_\tau(\varepsilon_i)\bigg)\eX_i , \quad i=1, \cdots , m+T_m,\\
\rr_{m,k}(\eu) & \equiv \sum^{m+k}_{i=m+1} \eR_i(\eu), \quad k=1, \cdots , T_m,
\end{align*} 
with $\eu \in \R^p$, $\| \eu \|_2 \leq C < \infty$. \\

By the following lemma we study the difference between $\rr_{m,k}(\eu)$ and its expectation, result which will be used to show the convergence in law under hypothesis $H_0$ of the two test statistics proposed in the following section.
\begin{lemma}
	\label{Lemma 4.2 (SQA)}
	For any constants $C_1> 0$,  $C_2 >0$ and integer $k \in \N$, we have that there exists a  constant  $ C_3 >0$ such that, for sufficiently large $k$ and $m$:
	\[
	\PP \bigg[\sup_{\| \eu \|_2 \leq C_1} \|\rr_{m,k}(\eu)  - \eE[\rr_{m,k}(\eu)] \|_1 \geq  2^{1/2} C_2 C_3 p m^{-1/2} k^{1/2}  ( \log k )^{1/2} \bigg] \leq 2 k^{-C_2^2}.
	\]	
\end{lemma}
The proof of Lemma \ref{Lemma 4.2 (SQA)} is given in Section \ref{sect_proof}. Underline   that Lemma \ref{Lemma 4.2 (SQA)} is true whether $H_0$ or $H_1$ hold. \\

With these preliminary elements we are now ready to introduce the test statistics and to state the main results of the present paper.
\section{Test statistics} 
\label{sect_results}
In this section we propose two test statistics, based on the expecile  and on the adaptive LASSO expectile residuals. For these statistics, we give the asymptotic distribution under  hypothesis $H_0$, for the two procedures (open-end and closed-end) and we show that they diverge under  hypothesis $H_1$. These results will allow us to find the stopping time when hypothesis $H_0$ starts to be rejected. \\
The proofs of all results presented in this section are given in Section \ref{sect_proof}.
\subsection{Test statistic based on the expectile residuals}
 In order to test  hypothesis $H_0$ against $H_1$, we first build a test statistic  based on the residuals obtained by considering the expectile estimator $\widehat{\eb}_m$  given by (\ref{hbm}) for $\eb_i$, with $i=m+1 , \cdots , m+T_m$. For this purpose, we first calculate the residuals on all data: 
\[\widehat{\varepsilon}_i =Y_i -\eX_i^\top \widehat{\eb}_m, \quad  \textrm{for } i=1, \cdots , m+T_m.
\]
After which, we  introduce:
 \begin{itemize}
\item the $p$-squared matrix: $\displaystyle{\eJ_m  \equiv \Var\![g_\tau(\varepsilon)] \frac{1}{m}\sum^m_{i=1} \eX_i \eX_i^\top= \Var\![g_\tau(\varepsilon)] \eO_m}$;
\item for  $\gamma \in [0,1/2)$ a fixed constant and $k=1, \cdots , T_m$, the following  normalization function 
\[z(m,k,\gamma)  \equiv m^{1/2} (1+k/m)(k/(k+m))^\gamma;
\]
\item the statistic $\displaystyle{\Gamma(m,k,\gamma)   \equiv  \frac{\| \eJ_m^{-1/2} \sum^{m+k}_{i=m+1} g_\tau(\widehat{\varepsilon}_i ) \eX_i \|_\infty}{z(m,k,\gamma)}}$.
\end{itemize}
Then, we consider as statistic for testing $H_0$ against $H_1$: 
$$ \sup_{1\leqslant k \leqslant T_m} \Gamma(m,k,\gamma), $$
which is a CUSUM (cumulative sum) test statistic calculated on the basis of the sum of $g_\tau(\widehat{\varepsilon}_i)$ weighted by the design and divided by the normalization function. By the following theorem, we state the asymptotic distribution of this test statistic under hypothesis $H_0$ for the open-end and closed-end procedures. Recall that $T=\lim_{m \rightarrow \infty}T_m/m$ for the closed-end procedure and for the open-end procedure we have $T=\infty$ .
\begin{theorem}
	\label{Theorem 2.1(SQA)}
	Under assumptions (A1)-(A3), if  hypothesis $H_0$ holds, then:\\
	\[
\sup_{1\leqslant k \leqslant T_m} \Gamma(m,k,\gamma)  \overset{\cal L} {\underset{m \rightarrow \infty}{\longrightarrow}} \sup_{0 < t < L(T) } \frac{\| \textbf{W}_p(t)\|_\infty}{t^\gamma}  ,
	\] 
with $\{\textbf{W}_p(t); \; t \in (0,\infty)\}$ a Wiener process of dimension $p$, $L(T)=1$ for the open-end procedure case and $L(T)=T/(1+T)$ for the closed end procedure case.
\end{theorem}

As stated in Section \ref{Intro} of  Introduction, \cite{Zhang-Li.17} also used the expectile method to estimate the parameters of the model, except that the change does not occur in the coefficients but when one of the continuous explanatory variables  exceeds an unknown threshold. The test statistic is different from ours and its asymptotic distribution under hypothesis $H_0$ is also different from that obtained in Theorem \ref{Theorem 2.1(SQA)}.\\
 
We now study the asymptotic behavior of the test statistic $\sup_{1\leqslant k \leqslant T_m} \Gamma(m,k,\gamma)$ under alternative hypothesis $H_1$, by proving that in this case it diverges. \\

 By the following theorem we prove that the test statistic built on the expectile residuals converges to infinity under the alternative hypothesis.
\begin{theorem}
	\label{Theorem 2.2(SQA)}
	Under assumptions (A1)-(A4),  if for any constant  $C_i$ such that  $ |C_i| \in [2\underline{c}, 2 \bar{c}]$ for $i=1, \cdots , m^s$, the condition  $m^{-s} \big\| \sum^{m+k^0_m+m^s }_{i=m+k^0_m+1} C_i \eX_i \eX_i^\top  \big\|_\infty > C >0$ is satisfied, then, when hypothesis $H_1$ is true such that $m^{1/2}\| \eb^1- \eb^0\|_2  {\underset{m \rightarrow \infty}{\longrightarrow}} \infty$, we have:\\
	\[
\sup_{1\leqslant k \leqslant T_m} \Gamma(m,k,\gamma)  \overset{\PP} {\underset{m \rightarrow \infty}{\longrightarrow}} \infty .
	\] 
\end{theorem}

The condition $m^{1/2}\| \eb^1- \eb^0\|_2  {\underset{m \rightarrow \infty}{\longrightarrow}} \infty$ indicates that the jump in the coefficients must be much greater than the convergence rate of the expectile estimator towards $\eb^0$. Intuitively, if the jump is of the same order $m^{-1/2}$ as the convergence rate then it will be difficult to identifying the change. The same condition was considered by \cite{Zou-Wang-Tang.15}, \cite{Ciuperca-17} for real-time change-point detection by quantile frameworks. Obviously, the parameters $\eb^1$,  $\eb^0$ may not depend on $m$.    \\

Theorems \ref{Theorem 2.1(SQA)} and \ref{Theorem 2.2(SQA)} allow us to detect the observation, called \textit{stopping time}, where the change will occur when   hypothesis $H_0$ is rejected, for a fixed size $\alpha \in (0,1)$:
\begin{equation}
\label{km}
\widehat{k}_m \equiv \left\{
\begin{array}{l}
\inf \{ k \geq 1; \; \; \Gamma(m,k,\gamma) > c_{\alpha}(\gamma)    \} \\
\infty ,  \qquad \qquad \textrm{if } \Gamma(m,k,\gamma) \leq c_{\alpha}(\gamma), \textrm{ for all }  k=1, \cdots , T_m,
\end{array}
\right.
\end{equation}
with $c_{\alpha}(\gamma)$ the $(1- \alpha)$-th quantile of the distribution of $ \sup_{0 < t < L(T) }  {\| \textbf{W}_p(t)\|_\infty}/{t^\gamma}$. \\
From Theorems \ref{Theorem 2.1(SQA)} and \ref{Theorem 2.2(SQA)} we also deduce that the asymptotic critical region is:  $  \Gamma(m,k,\gamma) > c_{\alpha}(\gamma)$. Moreover, for a  size  $\alpha \in (0,1)$ fixed, the test statistic  $\sup_{1 \leq k \leq T_m} \Gamma(m,k,\gamma)$ has the asymptotic type I error probability equal to $\alpha$ and the asymptotic power equal to 1, when $m \rightarrow \infty$. This implies that $\lim_{m \rightarrow \infty} \PP[\widehat{k}_m < \infty | H_0 \textrm{ true} ]=\alpha$ and $\lim_{m \rightarrow \infty} \PP[\widehat{k}_m < \infty | H_1 \textrm{ true} ]=1$, that is, that the asymptotic probability of a false alarm is $\alpha$ and that the change-point is detected with probability converging to one. 

\subsection{Test statistics based on the adaptive LASSO expectile residuals}
Now consider for model (\ref{eq1}) the possibility that the  number $p$ of regressors is large and that among the $p$ components $\beta^0_j$ of $\eb^0$ a certain number are 0. Then, the automatic selection of the relevant explanatory variables (that is with non-nul coefficients) is essential and therefore the test statistic will take into account the automatic selection results. We then use as an estimation method, the adaptative LASSO expectile method, proposed and studied by \cite{Liao-Park-Choi.19} for $p$ fixed and generalized by \cite{Ciuperca.19} when $p$ depends on $m$. In the  present paper we suppose that $p$ does not depend on  $m$ and that $p \leq m$. The condition $p \leq m$ is necessary to have the identifiability of the expectile estimator $\widehat{\eb}_m$ calculated on the historical data. \\
The adaptative LASSO expectile estimator calculated on the  historical observations is defined by:
\begin{equation}
\label{def_eb*}
\widehat{\eb}^*_m \equiv \argmin_{\eb \in \R^p} \bigg( \sum^m_{i=1} \rho_\tau(Y_i -\eX^\top_i \eb) +m \lambda_m \sum^p_{j=1} \widehat{\omega}_{m,j}|\beta_j| \bigg),
\end{equation}
with the adaptive weight $\widehat{\omega}_{m,j} \equiv | \widehat{\beta}_{m,j}|^{-\varphi}$, where $\widehat{\beta}_{m,j}$ is the $j$-th component of the expectile estimator $\widehat{\eb}_m$,  $\varphi >0$ a known constant and  $(\lambda_m)_{m \in \N}$ a positive deterministic sequence of tuning which converges to infinity as $m \rightarrow \infty$. The components of the adaptive LASSO  expectile estimator $\widehat{\eb}^*_m$ are $( \widehat{\beta}^*_{m,1}, \cdots , \widehat{\beta}^*_{m,p})^\top$.\\

The tuning parameter $\lambda_m $ and the  constant $\varphi$ satisfy the following assumption:
\begin{description}
	\item[\textbf{(A5)}] $m^{1/2} \lambda_m \rightarrow 0$ and $m^{(\varphi+1)/2} \lambda_m \rightarrow \infty$, for  $m \rightarrow \infty$.
\end{description}
\vspace{0.25cm}
Assumption (A5) is the classical supposition on the tuning parameter and on the power $\varphi$ for the adaptive LASSO estimators: see \cite{Zou.06} for the  LS loss function, \cite{Ciuperca-16} for the quantile loss function, \cite{Liao-Park-Choi.19} for the expectile loss function.\\

Let us consider the   set ${\cal A}^0 \equiv \{j \in \{1, \cdots , p\}; \;  \beta^0_j \neq 0 \} $ which contains the index of the non-zero coefficients   of model (\ref{eq1}). Since $\eb^0$ is unknown, the set ${\cal A}^0$ is unknown as well. Taking into account definition (\ref{def_eb*}) of the adaptive LASSO expectile estimator $\widehat{\eb}^*_m$, we deduce that the set  $ \widehat{\cal A}^*_m \equiv \{j \in \{1, \cdots , p\}; \; \widehat{\beta}^*_{m,j} \neq 0 \}$ is an estimator for ${\cal A}^0$.\\
For a $p$-vector $\eb=(\beta_1, \cdots , \beta_p )$ of parameters, let us denote by $\eb_{{\cal A}^0}$ the sub-vector of $\eb$ which contains the components  $\beta_j$, with $j \in {\cal A}^0$ and we denote by $\eX_{i,{\cal A}^0}=(X_{ij})_{j \in {\cal A}^0}$. We also define the $|{\cal A}^0|$-squared matrix  $\eO_{{\cal A}^0}$ by taking the lines and columns  $j \in {\cal A}^0$ of the matrix $\eO$. Thus, we have: $\eO_{{\cal A}^0}=\lim_{m \rightarrow \infty} \sum^m_{i=1} \eX_{i,{\cal A}^0} \eX^\top_{i,{\cal A}^0}$. Similarly, we define the vectors  $\eb_{\widehat{\cal A}^*_m}$, $\eX_{i,\widehat{\cal A}^*_m}$ and the matrix $\eO_{m,\widehat{\cal A}^*_m}=\sum^m_{i=1} \eX_{i,\widehat{\cal A}^*_m} \eX^\top_{i,\widehat{\cal A}^*_m}$.  \\

The interest of the estimator $\widehat \eb_m^*$  is that it satisfies \textit{oracle properties}, that is, under assumptions (A1)-(A3), (A5),  it satisfies the following two properties (see Theorem 3 of \cite{Liao-Park-Choi.19}):
\begin{itemize}
	\item \textit{sparsity property}: $\lim_{m \rightarrow \infty} \PP\big[{\cal A}^0 = \widehat{{\cal A}}^*_m \big]=1$.
	\item \textit{asymptotic normality property}:    $m^{1/2}  ( \widehat{\eb}_m^* - \eb^0)_{{\cal A}^0} $ converges in distribution to a zero-mean   Gaussian vector with the variance $\mu^{-2}_h\Var \![g_\tau(\varepsilon)] \eO^{-1}_{{\cal A}^0}$.  
\end{itemize}

The convergence rate of $\widehat \eb_m^*$ towards $\eb^0$ is of order $m^{-1/2}$, under assumptions (A1)-(A3) and the sequence $(\lambda_m)_{m \in \N }$ such that $m^{1/2} \lambda_m {\underset{m \rightarrow \infty}{\longrightarrow}} 0$ (see Theorem 2.1 of \cite{Ciuperca.19}).  \\

Let us now give a  similar result to Proposition \ref{Propositon 1} for the adaptive LASSO expectile estimators of the non-zero coefficients.
\begin{proposition}
	\label{Propositon 1A}
	Under assumptions (A1)-(A3), (A5), we have:
	\[
	\widehat{\eb}^*_{m, {\cal A}^0} =\eb^0_{{\cal A}^0} + \mu_{h }^{-1} \eO^{-1}_{{\cal A}^0} \frac{1}{m} \sum^{m}_{i=1} g_\tau(\varepsilon_i) \eX_{i,{\cal A}^0} +o_{\PP}(m^{-1/2}).
	\]
\end{proposition}
The proof of Proposition  \ref{Propositon 1A} is given in Section \ref{sect_proof}.\\
 
After this presentation of the necessary tools, we return to the test of $H_0$ against $H_1$. For $\gamma \in [0, 1/2)$ a fixed known constant and $k=1, \cdots , T_m$, we propose     in the case of a model with large number of explanatory variables the following  statistic:
\[
\Gamma^*(m,k,\gamma) \equiv \frac{\bigg\|\eJ^{-1/2}_{m, \widehat{\cal A}^*_m} \sum^{m+k}_{i=m+1}g_\tau(\widehat{\varepsilon}^*_i)  \eX_{i, \widehat{\cal A}^*_m}  \bigg\|_\infty}{z(m,k,\gamma)},
\]
with 
\[\widehat{\varepsilon}^*_i \equiv Y_i - \eX_i^\top \widehat{\eb}^*_m = Y_i -\eX_{i,\widehat{\cal A}^*_m}^\top \widehat{\eb}^*_{m, \widehat{\cal A}^*_m} \]
  the residuals corresponding to the adaptive LASSO expectile estimator and $\eJ_{m, \widehat{\cal A}^*_m}=\Var \![g_\tau(\varepsilon)] \eO_{m,\widehat{\cal A}^*_m}$. Thus, in order to test $H_0$ against $H_1$ when $p$ is large we will consider the following test statistic: 
\[	\sup_{1 \leq k \leq T_m}  \Gamma^*(m,k,\gamma).\]
For simplification of the calculations, taking into account the sparsity property of the estimator $\widehat{\eb}_m^*$, without reducing the generality of our approach, let us assume bellow that   $\widehat{\cal A}^*_m \subseteq {\cal A}^0 $,  other cases  are similarly proved. Then, we can write:
\begin{equation}
\label{eq25bis}
 \eX_{i, \widehat{\cal A}^*_m}^\top (\eb^0_{\widehat{\cal A}^*_m}+m^{-1/2} \eu^*_m)=\eX_{i,{\cal A}^0}^\top(\eb^0_{{\cal A}^0} +m^{-1/2}(\eu^*_m,\textbf{0}_{|{\cal A}^0 \cap \widehat{\cal A}^{*^c}_m|}))- \eX^\top_{i,{\cal A}^0 \cap \widehat{\cal A}^{*^c}_m} \eb^0_{{\cal A}^0 \cap \widehat{\cal A}^{*^c}_m},
\end{equation}
with $\eu^*_m \in \R^{|\widehat{\cal A}^*_m|}$, $\| \eu^*_m \|_2 \leq C$, the dimension of the vector $\eu^*_m$ being random.\\

 For studying the test statistic $	\sup_{1 \leq k \leq T_m}  \Gamma^*(m,k,\gamma)$ under hypothesis   $H_0$ we will first consider and study the following statistic:
 \[
 \Theta(m,k,\gamma)  \equiv \frac{\|\eJ^{-1/2}_{m,{\cal A}^0}  \sum^{m+k}_{i=m+1} g_\tau(\widehat{\varepsilon}^*_i) \eX_{i,{\cal A}^0} \|_\infty}{z(m,k,\gamma)} .
  \]
The following theorem gives the asymptotic distribution of the test statistic $\sup_{1 \leq k \leq T_m} \Theta(m,k,\gamma) $, under hypothesis $H_0$.
 \begin{theorem}
 	\label{Theorem 1(Metrika)}
 	Under assumptions (A1)-(A3), (A5),  if hypothesis $H_0$ holds, then:
 	\[
 	\sup_{1 \leq k \leq T_m} \Theta(m,k,\gamma) \overset{\cal L} {\underset{m \rightarrow \infty}{\longrightarrow}} \sup_{0 < t < L(T) } \frac{\|\textbf{W}_{|{\cal A}^0|}(t) \|_\infty}{t^\gamma},
 	\] 
 with $L(t)=1$ for the  open-end procedure, $L(T)=T/(1+T)$ for the closed-end procedure and    $\big\{ \textbf{W}_{|{\cal A}^0|}(t) , t \in (0, \infty )\big\}$  a  Wiener process of  dimension $|{\cal A}^0|$. 
 \end{theorem}
Then, we now state the asymptotic behaviour of the test statistic $\sup_{1 \leq k \leq T_m}  \Gamma^*(m,k,\gamma)$ under hypothesis $H_0$.  
\begin{corollary}
Under the same conditions as in  Theorem \ref{Theorem 1(Metrika)}, if hypothesis $H_0$ holds, we have:
\[
	\sup_{1 \leq k \leq T_m}  \Gamma^*(m,k,\gamma) - \sup_{0 < t < L(T) } \frac{\|\textbf{W}_{ |\widehat{\cal A}^*_m|}(t) \|_\infty}{t^\gamma} \overset{\PP} {\underset{m \rightarrow \infty}{\longrightarrow}} 0,
\]
with $\{\textbf{W}_{|\widehat{\cal A}^*_m|}(t); \; t \in (0, \infty)\}$ a Wiener process of  dimension $|\widehat{\cal A}^*_m|$.
\label{Corollary 1(Metrika)}
\end{corollary}
Note that under hypothesis $H_0$,   the asymptotic behaviour of the statistic  $\sup_{1 \leq k \leq T_m}    \Gamma^*(m,k,\gamma)$    is the similar as that of the test statistic considered by \cite{Ciuperca-18} for an  adaptive  LASSO  quantile model. The simulations presented in Section \ref{simus} will show that in the case   $p$ large, the test statistic based on the  adaptive  LASSO expectile estimator gives better results than that based on the adaptive LASSO  quantile estimators. \\ 

We now consider that hypothesis $H_1$ is true and that the unknown  parameter  vector   $\eb^1$    after  observation $k^0_m$ can contain non-zero coefficients for indices other than those of $\eb^0$. Let be then the corresponding index set ${\cal A}^1 \equiv \{j \in \{1, \cdots, p \}; \; \beta^1_j \neq 0\}$ which is also unknown. Without loss the generality, we rewrite: $\eb^0=\big( {\eb^0_{{\cal A}^0}}^{\top}, {\eb^0_{{\cal A}^{0^c}}}^{\top}\big)^\top$, $\eb^1=\big( {\eb^1_{{\cal A}^1}}^{\top}, {\eb^1_{{\cal A}^{1^c}}}^{\top}\big)^\top$. In order to study the test statistics  $\Theta(m,k,\gamma)$ and $\Gamma^*(m,k,\gamma)$, under hypothesis $H_1$, let us consider the index set   $A \equiv {\cal A}^0 \cup {\cal A}^1$, the parameters  $\widetilde{\eb^1} \equiv  \big({\eb^1_{{\cal A}^1}}^{\top}, {\textbf{0}_{{\cal A}^{1^c} \cap {\cal A}^0}}^{\top}\big)^\top$, $\widetilde{\eb^0} \equiv  \big( {\eb^0_{{\cal A}^0}}^{\top}, {\textbf{0}_{{\cal A}^{0^c} \cap {\cal A}^1}}^{\top}\big)^\top$ and  $\widetilde{\eu} =(\eu^\top,{\textbf{0}_{{\cal A}^{0^c} \cap {\cal A}^1}}^{\top} )^\top$ with  $\eu \in \R^{|{\cal A}^0|}$.
\begin{theorem}
	\label{Theorem 2(Metrika)}
	If assumptions (A1)-(A5) are satisfied and if for any constant $C_i$ such that  $ |C_i| \in [2\underline{c}, 2 \bar{c}]$ for $i=1, \cdots , m^s$ the following inequality takes place $m^{-s} \big\| \sum^{m^s}_{i=1} C_i \eX_{i,{\cal A}^0}  \eX_{i,A}^\top \big(\widetilde{\eb^1} - \widetilde{\eb^0}+m^{-1/2}\widetilde{\eu} \big) \big\|_\infty >C>0$,  when hypothesis $H_1$ holds  such that $m^{1/2}\| \eb^1- \eb^0\|_2 {\underset{m \rightarrow \infty}{\longrightarrow}} \infty$, then
	\[
	\sup_{1 \leq k \leq T_m} \Theta(m,k,\gamma)  \overset{\PP} {\underset{m \rightarrow \infty}{\longrightarrow}} \infty .
	\]
\end{theorem}
 This theorem allows us to deduce the divergence under hypothesis $H_1$ of the test statistic $\sup_{1 \leq k \leq T_m} \Gamma^*(m,k,\gamma)$. This test statistic  will be used in applications because the set ${\cal A}^0$ is unknown, being estimated by $\widehat{\cal A}^*_m$.
\begin{corollary}
	\label{Corollaire 2(Metrika)}
	Under the same conditions as in  Theorem \ref{Theorem 2(Metrika)}, we have
	\[
	\sup_{1 \leq k \leq T_m} \Gamma^*(m,k,\gamma)  \overset{\PP} {\underset{m \rightarrow \infty}{\longrightarrow}} \infty .
	\] 
\end{corollary}
Similar to definition (\ref{km}), we can find  the stopping time  from which hypothesis $H_0$ will be rejected  when the test statistic is $\sup_{1 \leq k \leq T_m} \Gamma^*(m,k,\gamma)$. Thus, for a fixed size $\alpha \in (0,1)$, the stopping time is: 
\begin{equation}
\label{km*}
\widehat{k}^*_m \equiv \left\{
\begin{array}{l}
\inf \{ k \geq 1; \; \; \Gamma^*(m,k,\gamma) > c_{m,\alpha}(\gamma)    \} \\
\infty ,  \qquad \qquad \textrm{if } \Gamma^*(m,k,\gamma) \leq c_{m,\alpha}(\gamma), \textrm{ for all }  k=1, \cdots , T_m,
\end{array}
\right.
\end{equation}
with $c_{m,\alpha}(\gamma)$ the $(1- \alpha)$-th quantile of the distribution of $ \sup_{0 < t < L(T) }  {\| \textbf{W}_{ |\widehat{\cal A}^*_m|}(t)\|_\infty}/{t^\gamma}$. The quantile $c_{m,\alpha}(\gamma)$ depends on $m$ by the dimension $|\widehat{\cal A}_m^*|$ of the Wiener process $\textbf{W}_{ |\widehat{\cal A}^*_m|}(t)$. Then,   the asymptotic critical region is: $ \Gamma^*(m,k,\gamma) > c_{m,\alpha}(\gamma)$. Similarly to the test statistic constructed on the basis of the expectile estimation method, the test statistic  $\sup_{1 \leq k \leq T_m} \Gamma^*(m,k,\gamma)$ has asymptotic size $\alpha$ and the asymptotic power equal to 1, when $m \rightarrow \infty$. Thus, we have also that $\lim_{m \rightarrow \infty} \PP[\widehat{k}^*_m < \infty | H_0 \textrm{ true} ]=\alpha$ and $\lim_{m \rightarrow \infty} \PP[\widehat{k}^*_m < \infty | H_1 \textrm{ true} ]=1$. \\

We conclude this subsection by pointing out that when the model contains irrelevant  variables, then the use of the test statistic $\sup_{1 \leq k \leq T_m} \Gamma^*(m,k,\gamma) $ is preferable   to $\sup_{1 \leq k \leq T_m} \Gamma(m,k,\gamma) $, especially when $p$ is large. This choice will be confirmed by the simulations presented in the following section.

\section{Simulation study and applications}
\label{simus}
In this section we first perform a numerical study to illustrate our theoretical results and to compare our test statistics with  those built on the adaptive LASSO quantile (\cite{Ciuperca-18}) and the adaptive LASSO LS (\cite{Ciuperca-15}) residuals. Afterwards, we present two applications on real data. 
\subsection{Simulation study}
For the simulation study presented in this section, we use the following R language packages: package \textit{SALES}  with function \textit{ernet} for the expectile regression, the package \textit{quantreg} with function \textit{rq} for quantile regression, the package \textit{"lqa"} with the  function \textit{lqa} for LS model with adaptive LASSO penalty.\\
 In all simulations we consider that the number of non-zero coefficients on the historical data is three. More precisely ${\cal A}^0=\{1, 2, 3\}$, with $\beta^0_1=2$, $\beta^0_2=2$, $\beta^0_3=1$. Under hypothesis $H_1$, the change occurs in $k^0=100$ and  only the first coefficient  changes: $\beta^1_1=-2$, $\beta^1_2=2$, $\beta^1_3=1$. Then ${\cal A}^1=\{1, 2, 3\}$ as well.  Regarding design $\eX_i=(X_{i1}, \cdots , X_{ip})^\top$ we consider two possibilities, for $i=1, \cdots , m+T_m$:
\begin{description}
	\item[D1]: $X_{ij} \sim {\cal N}(0,1)$, for any $j \in \{1, \cdots ,p\}\smallsetminus \{3,5,7,9\}$,
	\item[D2]:  $X_{ij} \sim \chi^2(1)+j^2/m$, for any $j \in \{1, \cdots ,p\}\smallsetminus \{3,5,7,9\}$, 
\end{description}	
 and  $X_{i3} \sim {\cal N}(2,1)$, $X_{i5} \sim {\cal N}(1,1)$, $X_{i7} \sim {\cal N}(-1,1)$, $X_{i9} \sim \big({\cal N}(1,1)\big)^2$, for D1 and D2.\\
 In all simulations, for adaptive LASSO expectile estimator (\ref{def_eb*}), we consider the following value for the tuning parameter $\lambda_m=m^{-2/5}$ and, based on the simulation conclusions of   \cite{Ciuperca.19}, we choose   the power $\varphi =1$ in the adaptive weights $\widehat{\omega}_{m,j}$. For the model errors $(\varepsilon_i)_{1 \leqslant i \leqslant m+T_m}$, two distributions are considered: one symmetric, more precisely the reduced centered normal distribution ${\cal N}(0,1)$ and one asymmetric of exponential distribution ${\cal E}\!xp(-1.5)$, that is with the density function $exp(-(x+1.5))\e1_{x>-1.5}$. In order to estimate  the expectile index  $\tau$ such that $ \eE[g_\tau(\varepsilon)]=0$ in assumption (A3), we use the following relation:
 \begin{equation*}
 \label{te}
 \tau=\frac{\eE\big[\varepsilon \e1_{\varepsilon<0} \big]}{\eE\big[\varepsilon(\e1_{\varepsilon<0}- \e1_{\varepsilon>0})  \big]}.
 \end{equation*}
Then,  we calculate the empirical estimation of $\tau$ by:
 \[
 \widehat \tau=\frac{m^{-1}\sum^m_{i=1} \varepsilon_i \e1_{\varepsilon_i <0}}{m^{-1} \big( \sum^m_{i=1} \varepsilon_i \e1_{\varepsilon_i <0}-\sum^m_{i=1} \varepsilon_i \e1_{\varepsilon_i >0}\big)}.
 \]
The same type of penalties are considered for the adaptive LASSO quantile and LS methods,  with the same tuning parameter $m^{-2/5}$ and the powers in the weights  $1.225$ and $9/40$, respectively. We consider the quantile method for the index 0.5, that is the  least absolute deviations loss.\\
For all studies, the theoretical test size is set  $\alpha =0.05$.
\subsubsection{Study and comparison of empirical sizes and powers}   
In  Tables \ref{L1b} and \ref{L2b} we present the empirical test sizes ($\widehat{\alpha}$)  for the open-end and closed-end procedures using the test statistics: $\sup_{1 \leqslant k \leqslant T_m} \Gamma(m, k,\gamma)$ built on the residuals obtained by the expectile estimation method, $\sup_{1 \leqslant k \leqslant T_m}  \Gamma^*(m, k,\gamma)$  built on the adaptive LASSO  expectile residuals and a modified adaptive  LASSO expectile statistic. The modified method consists of first selecting the relevant variables  (with the non-null  coefficient estimations) and reconsidering an expectile model only with these explanatory variables. The number of  historical observations is $m \in \{100, 200, 300, 500\}$.  The number of the Monte Carlo replications is  10000 when $m=100$, 5000 when $m \in \{200, 300\}$ and 2000 when $m=500$. The results presented in  Table \ref{L1b} are obtained when the design is D1, while in Table \ref{L2b} the design is of  type D2. The number $p$ of explanatory variables is set either three, so there are no non-zero coefficients, or ten, so there are seven zero coefficients.  From Tableau \ref{L1b} we observe that the results clearly improve for $m \geq 200$ when $\varepsilon \sim {\cal N}(0,1)$. On the other hand, for  $\varepsilon \sim {\cal E}\!xp(-1.5)$  when $p=10$, $m \in \{100,200 \}$, we get that the empirical test sizes $\widehat{\alpha}$ are strictly greater than 0.05 for closed-end procedure  by the three expectile estimation methods. For $m=300$, if $\gamma =0$, then we obtain $\widehat{\alpha} \leq 0.05 $ using the test statistics constructed on the residuals obtained by the adaptive LASSO and modified  methods (except the case $p=10$ for the closed-end procedure). For $m=500$, if $\gamma \in \{0, 0.15\}$ then  $\widehat{\alpha} \leq 0.05$ and generally, for $\gamma=0$ the results are better than for the other two values of $\gamma$. The results of Tableau \ref{L2b} are significantly better than in Table \ref{L1b}. For $p=10$, the smaller $m$ is  the less the results obtained by the expectile method are good compared to those obtained by the adaptive  LASSO expectile method. Conversely, when $p=|{\cal A}^0|=3$, that is when we have a small number of regressors and without irrelevant variables, the two test statistics corresponding to the two estimation methods give similar results. Whether in  Table \ref{L1b} or \ref{L2b}, we always obtain in all cases the empirical powers $\widehat{\pi}$ equal to 1. \\
In view of these results, because in the following we consider models with certain zero coefficients we only consider the test statistic  on the  adaptive LASSO expectile residuals.\\
In  Tables \ref{L3_L1} and \ref{L3_L2} we compare the results obtained by:
\begin{itemize}
	\item the  test statistic $\sup_{1 \leqslant k \leqslant T_m}  \Gamma^*(m, k,\gamma)$ constructed on adaptive LASSO  expectile residuals,
	\item the statistic proposed by  \cite{Ciuperca-18} using  adaptive LASSO quantile residuals,
	\item the CUSUM statistic, in the case of the    open-end procedure,  proposed by \cite{Ciuperca-15} constructed on the  adaptive  LASSO   LS residuals, 
\end{itemize}
  for  design $(\eX_i)_{1 \leqslant i \leqslant m+T_m}$ of type D1 in Table \ref{L3_L1} and of type D2 in Table \ref{L3_L2}. Since the test  statistic on the    adaptive LASSO LS residuals  systematically makes false change-point detections, in order to improve results, \cite{Ciuperca-15}   proposed a modified procedure. So, we did the same for adaptive LASSO expectile and quantile frameworks, for comparing.  We present the  empirical test sizes ($\widehat{\alpha}$) and powers ($\widehat{\pi}$) when the number of the explanatory variables is large: either 100 or 250. From Tables  \ref{L3_L1} and \ref{L3_L2} we deduce that by the  open-end procedure we obtain $\widehat \alpha \leq 0.06$ when $(\eX_i)_{1 \leqslant i \leqslant m+T_m} \equiv D1$ and $\widehat \alpha \leq 0.13$ when $(\eX_i)_{1 \leqslant i \leqslant m+T_m} \equiv D2$, by the  adaptive LASSO expectile and modified methods, for $\gamma \in \{0, 0.15\}$.  For open-end procedure, if $\gamma \in \{0, 0.15\}$, then the values of $\widehat{\alpha}$  obtained by the test statistic on the adaptive LASSO expectile residuals  are similar to those obtained by the test statistic   on the adaptive LASSO quantile residuals, except in the case $p=250$ when the test statistic on the adaptive LASSO  expectile residuals give better results. We have the same observation for the two modified procedures. Note that for the test statistics by the  quantile frameworks, when $(\eX_i)_{1 \leqslant i \leqslant m+T_m} \equiv D2$, we obtain that the values of $\widehat{\pi}$ are very far from 1, so, the tests do not detect the change in the model. The modified  adaptive LASSO LS test gives very good results ($\widehat{\alpha} \leq 0.05$). When hypothesis $H_1$ is true, our tests show their superiority  in the detection of $k^0=100$, since  we obtain $\widehat{\pi}=1$. For the two possible values of $p$, the  tests by the quantile frameworks don't always detect change when $\varepsilon \sim {\cal E}\!xp(-1.5)$. When $\varepsilon \sim {\cal E}\!xp(-1.5)$, $(\eX_i)_{1 \leqslant i \leqslant m+T_m} \equiv D1$, if $p$ is large, then the modified  adaptive LASSO LS test don't detect the change also in a proportion of maximum  $80\%$. \\
We observe that, when $p$ is large, the tests which give better results in terms  of the change-point detection or of less false alarm, by each framework, are: adaptive LASSO expectile, adaptive LASSO quantile and  modified adaptive LASSO LS, for the open-end procedure. We then make a comparison between these three methods, for open-end procedure, when $p=10$ and $|{\cal A}^0|=3$, so when there are zero and non-zero coefficients and the value of $p$ is small compared to $m=300$. The results are presented in  Table \ref{L4}. We observe that for $\gamma \in \{0, 0.15\}$ the empirical test sizes are all smaller than theoretical test size $\alpha =0.05$. Otherwise, when the design is D1, the corresponding test statistic to the modified adaptive LASSO LS   method does not detect the change-point, especially when the errors are exponential, therefore they have an asymmetric distribution. When the design is D2 and the  model errors are exponential, it is the test statistic on the adaptive LASSO quantile residuals which does not detect the change. Moreover, from  Table \ref{L4} we deduce that the test statistic  on the adaptive LASSO expectile residuals gives very good results when the model error distribution is asymmetric.
 \begin{table}
	\caption{\footnotesize Empirical test sizes ($\widehat{\alpha}$) obtained by the test statistics for  open-end  and closed-end  procedure, on the residuals corresponding  expectile, adaptive LASSO expectile, modified adaptive LASSO expectile  estimators, when   $|{\cal A}^0|=3$, $T_m=300$ and design $\eX=D1$.}
	\begin{center}
		{\scriptsize
			\begin{tabular}{|c|c|c|c|cc|cc|cc|}\hline  
				$m$	& estimation & &  & \multicolumn{6}{c|}{$\gamma $} \\ 
				\cline{5-10} 
				& method & procedure & $\varepsilon$  & \multicolumn{2}{c|}{0} & \multicolumn{2}{c|}{0.15} &   \multicolumn{2}{c|}{0.45}   \\ 
				&     &  &   & $p=3$ & $p=10$ & $p=3$ & $p=10$ & $p=3$ & $p=10$ \\
					\hline \hline
					500	& adaptive LASSO expectile  & open-end & ${\cal N}(0,1)$ & 0  & 0.005 & 0   & 0.01  & 0.04 & 0.21 \\ 
					&  &  & ${\cal E}\!xp(-1.5)$  & 0.002 & 0.001& 0.005 & 0.03  &  0.11 &  0.27  \\ 
					\cline{3-10} 
					&     & closed-end & ${\cal N}(0,1)$ & 0  & 0.01  & 0.004  & 0.04 & 0.04  & 0.22 \\ 
					&  &  & ${\cal E}\!xp(-1.5)$ & 0.005 & 0.03 & 0.01 & 0.07  &  0.11  &  0.29  \\  
					\cline{2-10} 
					&  modif adaptive LASSO expectile  & open-end & ${\cal N}(0,1)$ & 0 & 0.001  &    0&  0.006  & 0.04 & 0.16 \\ 
					&  &  & ${\cal E}\!xp(-1.5)$ & 0.002 & 0.004 &   0.005 & 0.01 &  0.11 &  0.22 \\   
					\cline{3-10} 
					&     & closed-end & ${\cal N}(0,1)$  & 0  & 0.007 & 0.004 & 0.02 & 0.04 & 0.18 \\
					&  &  & ${\cal E}\!xp(-1.5)$  & 0.005   & 0.01  & 0.01 & 0.03& 0.11&  0.23  \\  
					\cline{2-10} 
					&expectile  & open-end  & ${\cal N}(0,1)$ & 0 & 0.002& 0 & 0.01 & 0.04 &   0.37  \\
					&  &  & ${\cal E}\!xp(-1.5)$ & 0.002  & 0.008  &   0.005 & 0.02 &  0.11 &  0.35  \\
					\cline{3-10} 
					&     & closed-end & ${\cal N}(0,1)$ & 0  & 0.01 & 0.004 & 0.05 & 0.04 & 0.40 \\ 
					&  &  & ${\cal E}\!xp(-1.5)$  & 0.005  & 0.02 & 0.01 & 0.06 & 0.11 &  0.38  \\	\hline
			300	& adaptive LASSO expectile  & open-end & ${\cal N}(0,1)$  & 0.01 & 0.02 & 0.03  & 0.05 & 0.07 &  0.14 \\  
					&  &  & ${\cal E}\!xp(-1.5)$   & 0.03 & 0.04 & 0.05 & 0.07  &  0.14 &  0.22  \\  
					\cline{3-10} 
					&     & closed-end & ${\cal N}(0,1)$ & 0.05 &0.08  & 0.07 & 0.11 & 0.08  & 0.15 \\ 
					&  &  & ${\cal E}\!xp(-1.5)$ & 0.08  & 0.10 & 0.10 & 0.13 &  0.15 &  0.23  \\ 
					\cline{2-10} 
					&  modif adaptive LASSO expectile   & open-end & ${\cal N}(0,1)$ & 0.002 & 0.004 & 0.01 & 0.01 & 0.03 & 0.07 \\   
					&  &  & ${\cal E}\!xp(-1.5)$  & 0.007 & 0.01   & 0.02  & 0.02  &  0.09 &  0.14  \\  
					\cline{3-10} 
					&     & closed-end & ${\cal N}(0,1)$ & 0.02  & 0.02 & 0.02 & 0.03  & 0.04 & 0.08 \\  
					&  &  & ${\cal E}\!xp(-1.5)$  & 0.03  & 0.03 & 0.04  & 0.05 &  0.10 &  0.14  \\ 
					\cline{2-10} 
					&expectile  & open-end  & ${\cal N}(0,1)$ & 0.002  & 0.02 & 0.01 & 0.05  & 0.03 &  0.32  \\
					&  &  & ${\cal E}\!xp(-1.5)$  & 0.007  & 0.04  & 0.02 & 0.10 &  0.09 &  0.40  \\
					\cline{3-10} 
					&     & closed-end & ${\cal N}(0,1)$ & 0.02 & 0.08 & 0.02 & 0.14 & 0.03 & 0.36 \\ 
					&  &  & ${\cal E}\!xp(-1.5)$ & 0.03  & 0.13 & 0.04 & 0.19 &  0.10 &  0.42  \\	\hline
					
			200	& adaptive LASSO expectile  & open-end & ${\cal N}(0,1)$  & 0.05 & 0.05 & 0.08  & 0.08 & 0.15 &  0.17 \\  
					&  &  & ${\cal E}\!xp(-1.5)$   & 0.08 & 0.08 & 0.12 & 0.13  &  0.25 &  0.27  \\  
					\cline{3-10} 
					&     & closed-end & ${\cal N}(0,1)$ & 0.13  & 0.15 & 0.14 & 0.17  & 0.16 & 0.19 \\ 
					&  &  & ${\cal E}\!xp(-1.5)$ & 0.17  & 0.19 & 0.20 & 0.22 &  0.26 &  0.28  \\ 
					\cline{2-10} 
					&  modif adaptive LASSO expectile   & open-end & ${\cal N}(0,1)$ & 0.01 & 0.02 & 0.02 & 0.05 & 0.08 & 0.13 \\   
					&  &  & ${\cal E}\!xp(-1.5)$  & 0.02 & 0.05   & 0.04  & 0.08  &  0.16 &  0.21  \\  
					\cline{3-10} 
					&     & closed-end & ${\cal N}(0,1)$ & 0.04  & 0.09 & 0.06 & 0.10  & 0.09 & 0.14 \\  
					&  &  & ${\cal E}\!xp(-1.5)$  & 0.06  & 0.12 & 0.08  & 0.14 &  0.17 &  0.22  \\ 
					\cline{2-10} 
					&expectile  & open-end  & ${\cal N}(0,1)$ & 0.01  & 0.08 & 0.02 & 0.14  & 0.08 &  0.33  \\
					&  &  & ${\cal E}\!xp(-1.5)$  & 0.02  & 0.12  & 0.04 & 0.20 &  0.16 &  0.44  \\
					\cline{3-10} 
					&     & closed-end & ${\cal N}(0,1)$ & 0.04 & 0.24 & 0.06 & 0.28 & 0.09 & 0.37 \\ 
					&  &  & ${\cal E}\!xp(-1.5)$ & 0.06  & 0.29 & 0.08 & 0.34 &  0.17 &  0.46  \\	\hline
			100	& adaptive LASSO expectile & open-end & ${\cal N}(0,1)$ & 0.07  & 0.14 & 0.09 & 0.16  & 0.10 & 0.17\\  
					&  &  & ${\cal E}\!xp(-1.5)$ & 0.16 & 0.20 & 0.19 & 0.24  &  0.23 &  0.31  \\  
					\cline{3-10} 
					&     & closed-end & ${\cal N}(0,1)$  & 0.19 & 0.27 & 0.18 & 0.26 & 0.11 & 0.19 \\  
					&  &  & ${\cal E}\!xp(-1.5)$  & 0.27 & 0.33 & 0.22 & 0.33 &  0.24 &  0.32  \\ 
					\cline{2-10} 
					&  modif  adaptive LASSO expectile  & open-end & ${\cal N}(0,1)$ & 0.03 & 0.05  & 0.04  & 0.07  & 0.05 & 0.08 \\
					&  &  & ${\cal E}\!xp(-1.5)$  & 0.06  & 0.09   & 0.08  & 0.12  &  0.15 &  0.20  \\ 
					\cline{3-10} 
					&     & closed-end & ${\cal N}(0,1)$ & 0.09 & 0.14  & 0.09 & 0.13 & 0.06 & 0.09 \\  
					&  &  & ${\cal E}\!xp(-1.5)$ & 0.13 & 0.19  & 0.14 & 0.19 &  0.16 &  0.21  \\   
					\cline{2-10} 
					&expectile  & open-end  & ${\cal N}(0,1)$  & 0.03 & 0.25 & 0.04 & 0.31   & 0.05 & 0.47  \\
					&  &  &  ${\cal E}\!xp(-1.5)$ & 0.06 & 0.31 & 0.08  & 0.37 &  0.15 &  0.51  \\
					\cline{3-10} 
					&     & closed-end & ${\cal N}(0,1)$ & 0.09 & 0.49 & 0.09 & 0.48 & 0.06 & 0.50 \\ 
					&  &  & ${\cal E}\!xp(-1.5)$ & 0.13 & 0.52 & 0.14 & 0.52 &  0.16 &  0.53  \\	\hline
				\end{tabular} 
			}
		\end{center}
		\label{L1b} 
	\end{table}

\begin{table}
	\caption{\footnotesize Empirical test sizes ($\widehat{\alpha}$) obtained by the test statistics for  open-end  and closed-end  procedure, on the residuals corresponding  expectile, adaptive LASSO expectile, modified adaptive LASSO expectile estimators, when   $|{\cal A}^0|=3$, $T_m=300$ and design $\eX=D2$.}
	\begin{center}
		{\scriptsize
			\begin{tabular}{|c|c|c|c|cc|cc|cc|}\hline  
				$m$	& estimation & &  & \multicolumn{6}{c|}{$\gamma $} \\ 
				\cline{5-10} 
				& method & procedure & $\varepsilon$  & \multicolumn{2}{c|}{0} & \multicolumn{2}{c|}{0.15} &   \multicolumn{2}{c|}{0.45}   \\ 
				&     &  &   & $p=3$ & $p=10$ & $p=3$ & $p=10$ & $p=3$ & $p=10$ \\
				\hline \hline
				500	& adaptive LASSO expectile  & open-end & ${\cal N}(0,1)$ & 0.005 & 0.001 & 0.01  & 0.003  & 0.24 & 0.16 \\ 
				&  &  & ${\cal E}\!xp(-1.5)$  & 0.02 & 0.003 & 0.06 & 0.02  &  0.30 &  0.20  \\ 
				\cline{3-10} 
				&     & closed-end & ${\cal N}(0,1)$ & 0.02  & 0.005  & 0.04  & 0.01 & 0.26  & 0.17 \\ 
				&  &  & ${\cal E}\!xp(-1.5)$ & 0.05 & 0.02 & 0.10 & 0.04  &  0.31  &  0.20  \\  
				\cline{2-10} 
				&  modif  adaptive LASSO expectile  & open-end & ${\cal N}(0,1)$ & 0.004 & 0.001 &    0.01 &  0.003  & 0.22 & 0.16 \\ 
				&  &  & ${\cal E}\!xp(-1.5)$ & 0.01 & 0.003 &   0.03 & 0.02 &  0.25 &  0.20 \\   
				\cline{3-10} 
				&     & closed-end & ${\cal N}(0,1)$  & 0.009  & 0.005 & 0.02 & 0.01 & 0.24 & 0.17 \\
				&  &  & ${\cal E}\!xp(-1.5)$  & 0.02   & 0.02  & 0.06 & 0.04& 0.26 &  0.20  \\  
				\cline{2-10} 
				&expectile  & open-end & ${\cal N}(0,1)$ & 0.004 & 0.03&  0.01 & 0.08 & 0.22 &   0.32 \\
				&  &  & ${\cal E}\!xp(-1.5)$ & 0.01  & 0.05  &   0.03 & 0.13 &  0.25 &  0.51  \\
				\cline{3-10} 
				&     & closed-end & ${\cal N}(0,1)$ & 0.009  & 0.09 & 0.02 & 0.13 & 0.24 & 0.54 \\ 
				&  &  & ${\cal E}\!xp(-1.5)$  & 0.02  & 0.12 & 0.06 & 0.20 & 0.26 &  0.52  \\	\hline 
	300	& adaptive LASSO expectile  & open-end & ${\cal N}(0,1)$  & 0.008 & 0.007 & 0.02  & 0.02 & 0.19 &  0.18 \\  
				&  &  & ${\cal E}\!xp(-1.5)$   & 0.04 & 0.04 & 0.08 & 0.08  &  0.30 &  0.27  \\  
				\cline{3-10} 
				&     & closed-end & ${\cal N}(0,1)$ & 0.03  & 0.03 & 0.05  & 0.04  & 0.20 & 0.19 \\ 
				&  &  & ${\cal E}\!xp(-1.5)$ & 0.09  & 0.09 & 0.13 & 0.12 &  0.31 &  0.28  \\ 
				\cline{2-10} 
				&  modif adaptive LASSO expectile   & open-end & ${\cal N}(0,1)$ & 0.004 & 0.004 & 0.01 & 0.001 & 0.17 & 0.17 \\   
				&  &  & ${\cal E}\!xp(-1.5)$  & 0.01 & 0.02   & 0.04  & 0.04  &  0.23 &  0.21  \\  
				\cline{3-10} 
				&     & closed-end & ${\cal N}(0,1)$ & 0.02  & 0.02 & 0.04 & 0.02  & 0.18 & 0.18 \\  
				&  &  & ${\cal E}\!xp(-1.5)$  & 0.04  & 0.05 & 0.07  & 0.07 &  0.24 &  0.22  \\ 
				\cline{2-10} 
				&expectile  & open-end  & ${\cal N}(0,1)$ & 0.004  & 0.02  & 0.01 & 0.07  & 0.17 &  0.14  \\
				&  &  & ${\cal E}\!xp(-1.5)$  & 0.01  & 0.07  & 0.04 & 0.16 &  0.23 &  0.48  \\
				\cline{3-10} 
				&     & closed-end & ${\cal N}(0,1)$ & 0.02 & 0.10 & 0.04 & 0.16 & 0.18 & 0.47 \\ 
				&  &  & ${\cal E}\!xp(-1.5)$ & 0.04  & 0.17 & 0.07 & 0.25 &  0.24 &  0.50  \\	\hline
	200	& adaptive LASSO expectile  & open-end & ${\cal N}(0,1)$  & 0.03 & 0.01 & 0.06  & 0.01 & 0.28 &  0.18 \\  
		&  &  & ${\cal E}\!xp(-1.5)$   & 0.10 & 0.05 & 0.17 & 0.08  &  0.39 &  0.27  \\  
		\cline{3-10} 
		&     & closed-end & ${\cal N}(0,1)$ & 0.10  & 0.04 & 0.13  & 0.04  & 0.30 & 0.19 \\ 
		&  &  & ${\cal E}\!xp(-1.5)$ & 0.19  & 0.10 & 0.25 & 0.12 &  0.40 &  0.28  \\ 
		\cline{2-10} 
		&  modif  adaptive LASSO expectile  & open-end & ${\cal N}(0,1)$ & 0.02 & 0.004 & 0.05 & 0.009 & 0.26 & 0.17 \\   
		&  &  & ${\cal E}\!xp(-1.5)$  & 0.05 & 0.02   & 0.11  & 0.05  &  0.31 &  0.22  \\  
		\cline{3-10} 
		&     & closed-end & ${\cal N}(0,1)$ & 0.07  & 0.02 & 0.11 & 0.02  & 0.27 & 0.18 \\  
		&  &  & ${\cal E}\!xp(-1.5)$  & 0.11  & 0.05 & 0.16  & 0.07 &  0.32 &  0.22  \\ 
		\cline{2-10} 
		&expectile  & open-end  & ${\cal N}(0,1)$ & 0.02  & 0.09  & 0.05 & 0.15  & 0.26 &  0.48  \\
		&  &  & ${\cal E}\!xp(-1.5)$  & 0.05  & 0.16  & 0.11 & 0.25 &  0.31 &  0.53  \\
		\cline{3-10} 
		&     & closed-end & ${\cal N}(0,1)$ & 0.07 & 0.23 & 0.11 & 0.27 & 0.27 & 0.51 \\ 
		&  &  & ${\cal E}\!xp(-1.5)$ & 0.11  & 0.30 & 0.16 & 0.36 &  0.32 &  0.54  \\	\hline
	100	& adaptive LASSO expectile & open-end & ${\cal N}(0,1)$ & 0.12  & 0.11 & 0.21 & 0.14  & 0.48 & 0.23\\  
				&  &  & ${\cal E}\!xp(-1.5)$ & 0.23 & 0.22 & 0.30 & 0.27  &  0.44 &  0.38  \\  
				\cline{3-10} 
				&     & closed-end & ${\cal N}(0,1)$  & 0.22 & 0.23 & 0.29 & 0.23 & 0.49 & 0.25 \\  
				&  &  & ${\cal E}\!xp(-1.5)$  & 0.27 & 0.35 & 0.28 & 0.36 &  0.31 &  0.40  \\ 
				\cline{2-10} 
				& modif  adaptive LASSO expectile   & open-end & ${\cal N}(0,1)$ & 0.09 & 0.08  & 0.18  & 0.10  & 0.45 & 0.19 \\
				&  &  & ${\cal E}\!xp(-1.5)$  & 0.13  & 0.13   & 0.19  & 0.18  &  0.41 &  0.29  \\ 
				\cline{3-10} 
				&     & closed-end & ${\cal N}(0,1)$ & 0.18 & 0.18  & 0.25 & 0.18 & 0.46 & 0.21 \\  
				&  &  & ${\cal E}\!xp(-1.5)$ & 0.20 & 0.24  & 0.26 & 0.25 &  0.42 &  0.31  \\   
				\cline{2-10} 
				&expectile  & open-end  & ${\cal N}(0,1)$  & 0.09 & 0.38 & 0.18 & 0.44   & 0.45 & 0.59  \\
				&  &  &  ${\cal E}\!xp(-1.5)$ & 0.13 & 0.44 & 0.19  & 0.50 &  0.41 &  0.61  \\
				\cline{3-10} 
				&     & closed-end & ${\cal N}(0,1)$ & 0.18 & 0.60 & 0.25 & 0.60 & 0.46 & 0.62 \\ 
				&  &  & ${\cal E}\!xp(-1.5)$ & 0.20 & 0.62 & 0.26 & 0.63 &  0.42 &  0.63  \\	\hline
			\end{tabular} 
		}
	\end{center}
	\label{L2b} 
\end{table}

\begin{table}
	\caption{\footnotesize Empirical test sizes ($\widehat{\alpha}$) and powers ($\widehat{\pi}$)  for  open-end  and closed-end  procedure, for the test statistics built on   the residuals of the adaptive(\textit{ad.}) LASSO,  modified(\textit{modif}) adaptive LASSO for  expectile, quantile and LS  frameworks, when  $|{\cal A}^0|=3$, $k^0=100$, $T_m=300$, $m=300$ and design $\eX=D1$.}
	\begin{center}
		{\scriptsize
			\begin{tabular}{|c|c|c|c|cc|cc|cc|}\hline  
				$\widehat{\alpha}$ or $\widehat{\pi}$ &	& & estimation &     \multicolumn{6}{c|}{$\gamma $} \\ 
				\cline{5-10} 
				& procedure & $\varepsilon$ & method  & \multicolumn{2}{c|}{0} & \multicolumn{2}{c|}{0.15} &   \multicolumn{2}{c|}{0.45}   \\ 
				&     &  &   & $p=100$ & $p=250$ & $p=100$ & $p=250$ & $p=100$ & $p=250$ \\ \hline \hline
				$\widehat{\alpha}$  & open-end & ${\cal N}(0,1)$ & ad.LASSO expectile  &0.01 & 0.005& 0.02& 0.01 & 0.115 & 0.05  \\     
				&   &  	& modif ad.LASSO expect   & 0.002 & 0.03 & 0.01& 0.04 & 0.10  &0.10 \\  
				&   &  	& ad.LASSO quantile  & 0.003 & 0.005 & 0.01 &  0.005& 0.05 &0.09 \\   
				&   &  	& modif ad.LASSO quant   & 0.003 &0.07 & 0.01&0.12 & 0.05 &0.20 \\ 
				&   &  	& ad.LASSO LS  & 1  & 1 & 1 & 1  & 1 & 1 \\    
				&   &  	& modif ad.LASSO LS   & 0.002 &0 & 0.005 & 0.01& 0.02 & 0.04 \\  
				\cline{3-10} 
				&  & ${\cal E}\!xp(-1.5) $	& ad.LASSO expectile   &0.04 &0.01& 0.07 &0.03 & 0.24 &0.12  \\
				&   &  	& ad.LASSO expect modif  & 0.02& 0.01 & 0.04 &0.04 & 0.19 &0.16 \\ 
				&   &  	& ad.LASSO quantile  & 0.009 &0 & 0.02 & 0.03& 0.14 &0.10  \\  
				&   &  	& ad.LASSO quant modif  & 0.01 &0.20 & 0.02&0.25 & 0.13 &0.33  \\
				&   &  	& ad.LASSO LS  & 1  & 1 & 1  & 1  & 1 & 1  \\    
				&   &  	&  modif ad.LASSO LS  & 0.008 & 0.03& 0.01 & 0.03 & 0.06 &0.09  \\ 
				\cline{2-10} 
				& closed-end & ${\cal N}(0,1)$ 	& ad.LASSO expectile   &0.15& 0.05 & 0.15 &0.05 & 0.15 &0.07 \\    
				&   &  	&  modif ad.LASSO expect  & 0.08 &0.11 & 0.09 &0.11 & 0.14 & 0.12 \\  
				&   &  	& ad.LASSO quantile  & 0.09 & 0.08 & 0.09&0.08 & 0.08&0.12 \\   
				&   &  	&  modif ad.LASSO quant  & 0.09&0.24 & 0.09& 0.25 & 0.07 &0.23 \\ 
				\cline{3-10} 
				&  & ${\cal E}\!xp(-1.5) $	& ad.LASSO expectile   &0.22 & 0.14& 0.23&0.15 & 0.27 &0.13  \\
				&   &  	& modif ad.LASSO expect   &0.11 & 0.16 & 0.13 &0.17 & 0.20  &0.18 \\  
				&   &  	& ad.LASSO quantile  & 0.12 & 0.10 & 0.13 &0.09 & 0.18& 0.11  \\ 
				&   &  	& modif ad.LASSO quant   &0.12&0.41 & 0.13&0.40  & 0.18&0.33  \\  
				\cline{1-10} 
				$\widehat{\pi}$   & open-end & ${\cal N}(0,1)$ 	& ad.LASSO expectile   &1 & 1 & 1  &1 & 1 & 1 \\  
				&   &  	& modif ad.LASSO expect   &1 & 1 & 1  &1 & 1 & 1 \\     
				&   &  	& ad.LASSO quantile &1 & 1 & 1  &1 & 1 & 1  \\      
				&   &  	& modif ad.LASSO quant   &1 & 1 & 1  &1 & 1 & 1   \\ 
				&   &  	& ad.LASSO LS  & 1  & 1 & 1 & 1  & 1 & 1  \\    
				&   &  	& modif ad.LASSO LS   &0.86 & 0.85& 0.93 & 0.92 & 0.97 &0.97\\  
				\cline{3-10} 
				&  & ${\cal E}\!xp(-1.5) $	& ad.LASSO expectile   &1 & 1 & 1  &1 & 1 & 1 \\  
				&   &  	& modif ad.LASSO expect   &1 & 1 & 1  &1 & 1 & 1 \\   
				&   &  	& ad.LASSO quantile &1 & 1 & 1  &1 & 1 & 1 \\ 
				&   &  	& modif ad.LASSO quant   &1 & 1 & 1  &1 & 1 & 1 \\ 
				&   &  	& ad.LASSO LS  & 0.97  & 0.99 & 0.98 & 0.99 & 0.98 & 0.99  \\    
				&   &  	& modif ad.LASSO LS   & 0.98 &0.45 & 0.99 &0.59& 0.99 &0.73  \\  
				\cline{2-10} 
				& closed-end & ${\cal N}(0,1)$ 	& ad.LASSO expectile   & 1  & 1 & 1  &1 & 1 & 1 \\  
				&   &  	& modif ad.LASSO expect   & 1  & 1 & 1  &1 & 1 & 1 \\ 
				&   &  	& ad.LASSO quantile  & 1  & 1 & 1 & 1  & 1 & 1 \\ 
				&   &  	& modif ad.LASSO quant   & 1 &1 & 1 & 1& 1  &1 \\ 
				\cline{3-10} 
				&  & ${\cal E}\!xp(-1.5)$ 	& ad.LASSO expectile   & 1  & 1 & 1  &1 & 1 & 1 \\  
				&   &  	& modif ad.LASSO expect   & 1  & 1 & 1  &1 & 1 & 1 \\ 
				&   &  	& ad.LASSO quantile  & 1  & 1 & 1  &1 & 1 & 1 \\   
				&   &  	& modif ad.LASSO quant   & 1  & 1 & 1  &1 & 1 & 1 \\  \hline
			\end{tabular}  
		}
	\end{center}
	\label{L3_L1}  
\end{table}

\begin{table}
	\caption{\footnotesize Empirical test sizes ($\widehat{\alpha}$) and powers ($\widehat{\pi}$)  for  open-end  and closed-end  procedure, for the test statistics built on   the residuals of the  adaptive(\textit{ad.}) LASSO, modified(\textit{modif}) adaptive LASSO  for  expectile, quantile and LS  frameworks, when  $|{\cal A}^0|=3$, $k^0=100$, $T_m=300$, $m=300$ and design $\eX=D2$ .}
	\begin{center}
		{\scriptsize
			\begin{tabular}{|c|c|c|c|cc|cc|cc|}\hline  
		  $\widehat{\alpha}$ or $\widehat{\pi}$ &	& & estimation &     \multicolumn{6}{c|}{$\gamma $} \\ 
				\cline{5-10} 
			& procedure & $\varepsilon$ & method  & \multicolumn{2}{c|}{0} & \multicolumn{2}{c|}{0.15} &   \multicolumn{2}{c|}{0.45}   \\ 
				&     &  &   & $p=100$ & $p=250$ & $p=100$ & $p=250$ & $p=100$ & $p=250$ \\ \hline \hline
		 $\widehat{\alpha}$  & open-end & ${\cal N}(0,1)$ & ad.LASSO expectile  &0.02 & 0.01& 0.03& 0.03 & 0.25 & 0.07  \\     
	     &   &  	&  modif ad.LASSO expect  & 0.03 & 0.04 & 0.05& 0.06 & 0.28  &0.11 \\  
	     &   &  	& ad.LASSO quantile  & 0.001 & 0.002 & 0.01 &  0.007& 0.06 &0.09 \\   
	     &   &  	& modif ad.LASSO quant   & 0.03 &0.09 & 0.06&0.13 & 0.12 &0.24 \\ 
	      &   &  	& ad.LASSO LS  & 1  & 1 & 1 & 1  & 1 & 1 \\    
	        &   &  	& modif ad.LASSO LS   & 0.004 &0.001 & 0.005 & 0.004& 0.02 & 0.02 \\  
	     \cline{3-10} 
	        &  & ${\cal E}\!xp(-1.5) $	& ad.LASSO expectile   &0.06 &0.06& 0.12 &0.09 & 0.36 &0.21  \\
	        &   &  	& modif ad.LASSO expect   & 0.08& 0.09 & 0.13 &0.12 & 0.36 &0.23 \\ 
	        &   &  	& ad.LASSO quantile  & 0.02 &0.01 & 0.05 & 0.03& 0.35 &0.10  \\  
	        &   &  	& modif ad.LASSO quant   & 0.17 &0.22 & 0.22&0.27 & 0.33 &0.37  \\
	        &   &  	& ad.LASSO LS  & 1  & 1 & 1  & 1  & 1 & 1  \\    
	       &   &  	& modif ad.LASSO LS    & 0.007 & 0.006& 0.01 & 0.01 & 0.06 &0.05  \\ 
	     \cline{2-10} 
	        & closed-end & ${\cal N}(0,1)$ 	& ad.LASSO expectile   &0.16& 0.13 & 0.16 &0.14 & 0.29 &0.09 \\    
	       &   &  	& modif ad.LASSO expect   & 0.26 &0.18 & 0.28 &0.18 & 0.38 & 0.13 \\  
	       &   &  	& ad.LASSO quantile  & 0.12 & 0.11 & 0.12&0.11 & 0.09&0.10 \\   
	     &   &  	& modif ad.LASSO quant    & 0.21&0.33 & 0.21& 0.32 & 0.16 &0.27 \\ 
	    	     \cline{3-10} 
	       &  & ${\cal E}\!xp(-1.5)$ 	& ad.LASSO expectile   &0.23 & 0.21& 0.24&0.21 & 0.28 &0.23  \\
	         &   &  	& modif ad.LASSO expect   &0.26 & 0.25 & 0.29 &0.26 & 0.38  &0.26 \\  
	        &   &  	& ad.LASSO quantile  & 0.20 & 0.17 & 0.21 &0.17 & 0.37& 0.12  \\ 
	        &   &  	& modif ad.LASSO quant    &0.40&0.46 & 0.40&0.46  & 0.52&0.41  \\  
	      \cline{1-10} 
	    $\widehat{\pi}$   & open-end & ${\cal N}(0,1)$ 	& ad.LASSO expectile   &1 & 1 & 1  &1 & 1 & 1 \\  
	      &   &  	& modif ad.LASSO expect   &1 & 1 & 1  &1 & 1 & 1 \\     
	    &   &  	& ad.LASSO quantile  & 0.99 & 0.99 & 0.99 & 1& 1  & 1 \\      
	      &   &  	& modif ad.LASSO quant  &0.99 &0.98 & 0.99 &0.99 & 1 &  1  \\ 
	    &   &  	& ad.LASSO LS  & 1  & 1 & 1 & 1  & 1 & 1  \\    
	    &   &  	& modif ad.LASSO LS    & 1  & 1 & 1 & 1  & 1 & 1\\  
	   \cline{3-10} 
	      &  & ${\cal E}\!xp(-1.5)$ 	& ad.LASSO expectile   &1 & 1 & 1  &1 & 1 & 1 \\  
	     &   &  	& modif ad.LASSO expect   &1 & 1 & 1  &1 & 1 & 1 \\   
	     &   &  	& ad.LASSO quantile  & 0.42  &0.51& 0.55 & 0.65& 0.73 &0.79 \\ 
	      &   &  	& modif ad.LASSO quant   & 0.30&0.33 & 0.43& 0.49 & 0.63 &0.75 \\ 
	    &   &  	& ad.LASSO LS  & 1  & 1 & 1 & 1  & 1 & 1  \\    
	     &   &  	& modif ad.LASSO LS   & 1 &1 & 1 &1& 1 &1  \\  
	   \cline{2-10} 
	   & closed-end & ${\cal N}(0,1)$ 	& ad.LASSO expectile   & 1  & 1 & 1  &1 & 1 & 1 \\  
	     &   &  	& modif ad.LASSO expect   & 1  & 1 & 1  &1 & 1 & 1 \\ 
	   &   &  	& ad.LASSO quantile  & 1  & 1 & 1 & 1  & 1 & 1 \\ 
	    &   &  	& modif ad.LASSO quant   & 1  & 1 & 1  &1 & 1 & 1  \\ 
	   \cline{3-10} 
	    &  &$ {\cal E}\!xp(-1.5) $	& ad.LASSO expectile   & 1  & 1 & 1  &1 & 1 & 1 \\  
	     &   &  	& modif ad.LASSO expect   & 1  & 1 & 1  &1 & 1 & 1 \\ 
	    &   &  	& ad.LASSO quantile  & 0.87 &0.93& 0.86&0.93 & 0.72 &0.84 \\   
	    &   &  	& modif ad.LASSO quant   &0.79 &0.82 & 0.79& 0.84 & 0.69 &0.81 \\  \hline
	  \end{tabular}  
		}
	\end{center}
	\label{L3_L2}   
\end{table}

\begin{table}
	\caption{\footnotesize Empirical test sizes ($\widehat{\alpha}$) and powers ($\widehat{\pi}$)  for  open-end   procedure, for the test statistics on   the residuals of the adaptive(\textit{ad.}) LASSO for  expectile, quantile and  modified(\textit{modif})  LS  frameworks, when $|{\cal A}^0|=3$, $m=300$, $p=10$, $k^0=100$, $T_m=300$.}
	\begin{center}
		{\scriptsize
			\begin{tabular}{|c|c|c|c|c|c|c|}\hline  
			$\eX$ & 	$\widehat{\alpha}$ or $\widehat{\pi}$ &$\varepsilon$	& estimation &     \multicolumn{3}{c|}{$\gamma $} \\ 
				\cline{5-7} 
				&  &  & method  & 0 & 0.15 &  0.45  \\  \hline \hline
					D1 &	$\widehat{\alpha}$  &   ${\cal N}(0,1)$ & ad.LASSO expectile  &0.007 & 0.02& 0.07 \\       
				&   &  	& ad.LASSO quantile  & 0.004 & 0.008 & 0.02   \\   
				&   &  	& modif ad.LASSO LS   & 0.002 &0.004 & 0.02   \\  
				\cline{3-7} 
				&  & ${\cal E}\!xp(-1.5) $	& ad.LASSO expectile   &0.002 &0.04& 0.18   \\ 
				&   &  	& ad.LASSO quantile  & 0.007 &0.02 & 0.11    \\  
				&   &  	& modif ad.LASSO LS   & 0.01 & 0.02& 0.08   \\ 
				\cline{2-7} 
				&	$\widehat{\pi}$  &   ${\cal N}(0,1)$ & ad.LASSO expectile  &1 & 1& 1  \\       
				&   &  	& ad.LASSO quantile  &1 & 1& 1   \\   
				&   &  	& modif ad.LASSO LS   &0.48& 0.62& 0.73   \\  
				\cline{3-7} 
				&  & ${\cal E}\!xp(-1.5) $	& ad.LASSO expectile   &1 & 1& 1   \\ 
				&   &  	& ad.LASSO quantile  &1 & 1& 1    \\  
				&   &  	&  modif ad.LASSO LS  &0.08 & 0.16& 0.30   \\ 
				\cline{1-7} 				
			D2 &	$\widehat{\alpha}$  &   ${\cal N}(0,1)$ & ad.LASSO expectile  &0.002 & 0.008& 0.14  \\       
				&   &  	& ad.LASSO quantile  & 0.001 & 0.004 & 0.06   \\   
				&   &  	& modif ad.LASSO LS   & 0.003 &0.006 & 0.03   \\  
				\cline{3-7} 
				&  & ${\cal E}\!xp(-1.5) $	& ad.LASSO expectile   &0.02 &0.04& 0.22   \\ 
				&   &  	& ad.LASSO quantile  & 0.007 &0.02 & 0.18    \\  
				&   &  	& modif  ad.LASSO LS  & 0.01 & 0.02& 0.07   \\ 
				\cline{2-7} 
			 &	$\widehat{\pi}$  &   ${\cal N}(0,1)$ & ad.LASSO expectile  &1 & 1& 1  \\       
			&   &  	& ad.LASSO quantile  & 0.99 & 0.99 & 0.99   \\   
			&   &  	& modif ad.LASSO LS   &1 & 1& 1   \\  
			\cline{3-7} 
			&  & ${\cal E}\!xp(-1.5) $	& ad.LASSO expectile   &1 & 1& 1   \\ 
			&   &  	& ad.LASSO quantile  & 0.08 &0.17 & 0.43    \\  
			&   &  	& modif  ad.LASSO LS  &1 & 1& 1   \\ 
			\cline{1-7} 
		\end{tabular}  
		}
	\end{center}
	\label{L4}   
\end{table}
\subsubsection{Automatic variable selection comparison}
\begin{figure}[h!]
	\includegraphics[width=15cm,height=5cm,angle=0]{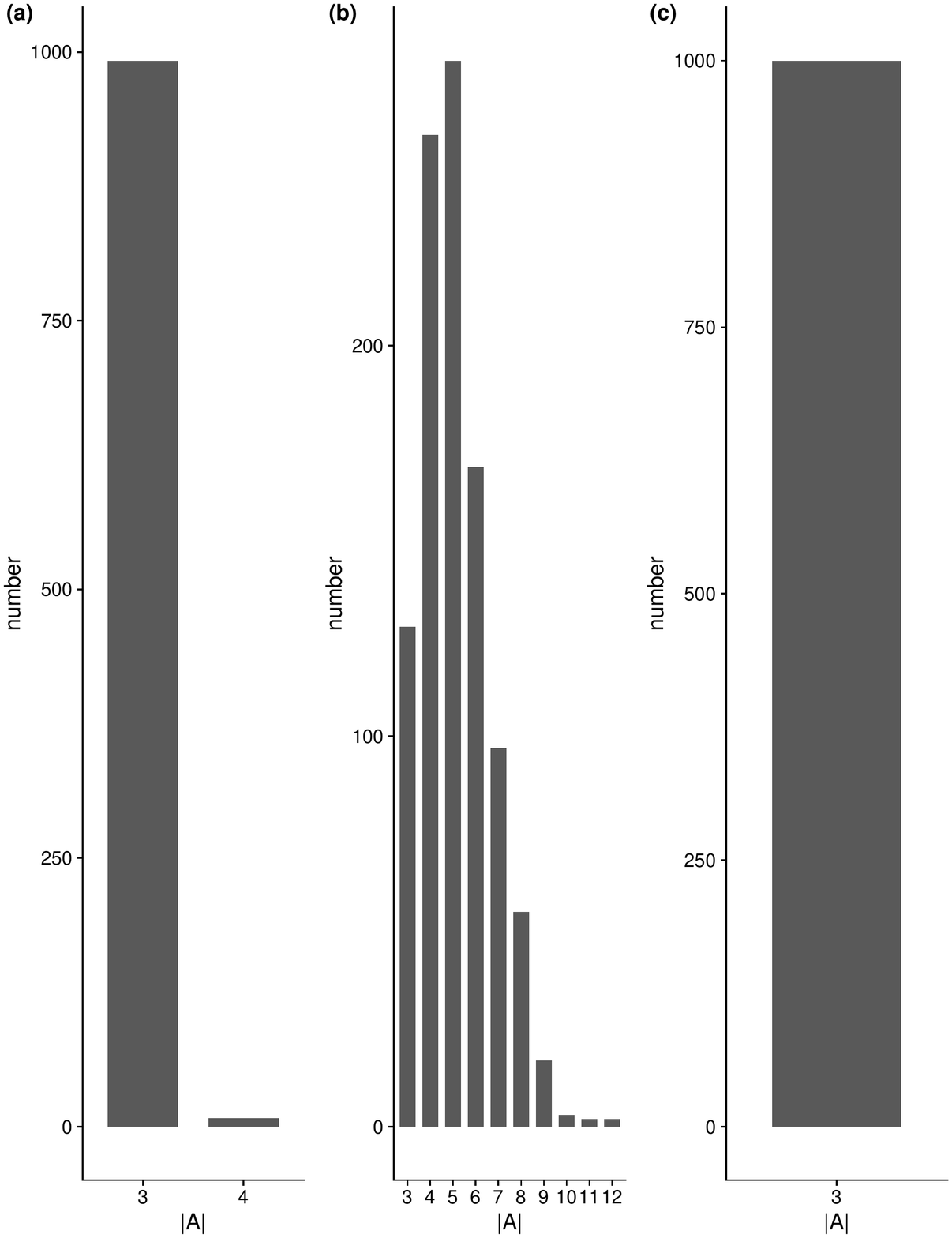}
	\caption{ Histograms for 1000 Monte Carlo replications of $|\widehat{\cal A}_m|$ for adaptive LASSO expectile (a), adaptive LASSO quantile(b) and adaptive LASSO  LS (c) estimations, when the design  $\eX$ is D1,   $m=300$, $p=250$, $|{\cal A}^0|=3$, $\varepsilon \sim{\cal N}(0,1)$.}
	\label{fig_A_N_L1} 
\end{figure}

\begin{figure}[h!]
	\includegraphics[width=15cm,height=5cm,angle=0]{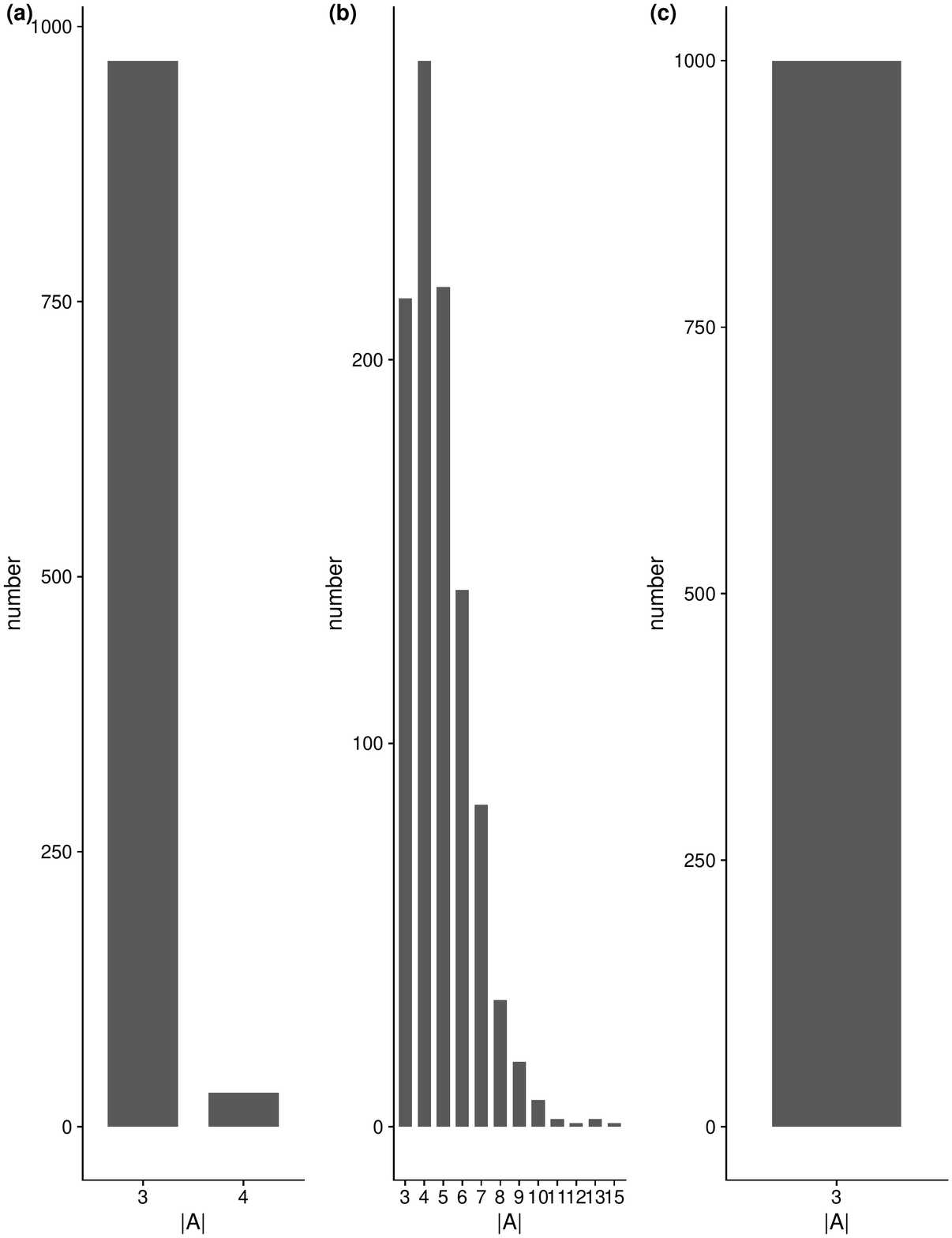}
	\caption{ Histograms for 1000 Monte Carlo replications of $|\widehat{\cal A}_m|$ for adaptive LASSO expectile (a), adaptive LASSO quantile(b) and adaptive LASSO  LS (c) estimations, when the design  $\eX$ is D1,    $m=300$, $p=250$, $|{\cal A}^0|=3$, $\varepsilon \sim{\cal E}\!xp(-1.5)$.}
	\label{fig_A_E_L1} 
\end{figure}

\begin{figure}[h!]
	\includegraphics[width=15cm,height=5cm,angle=0]{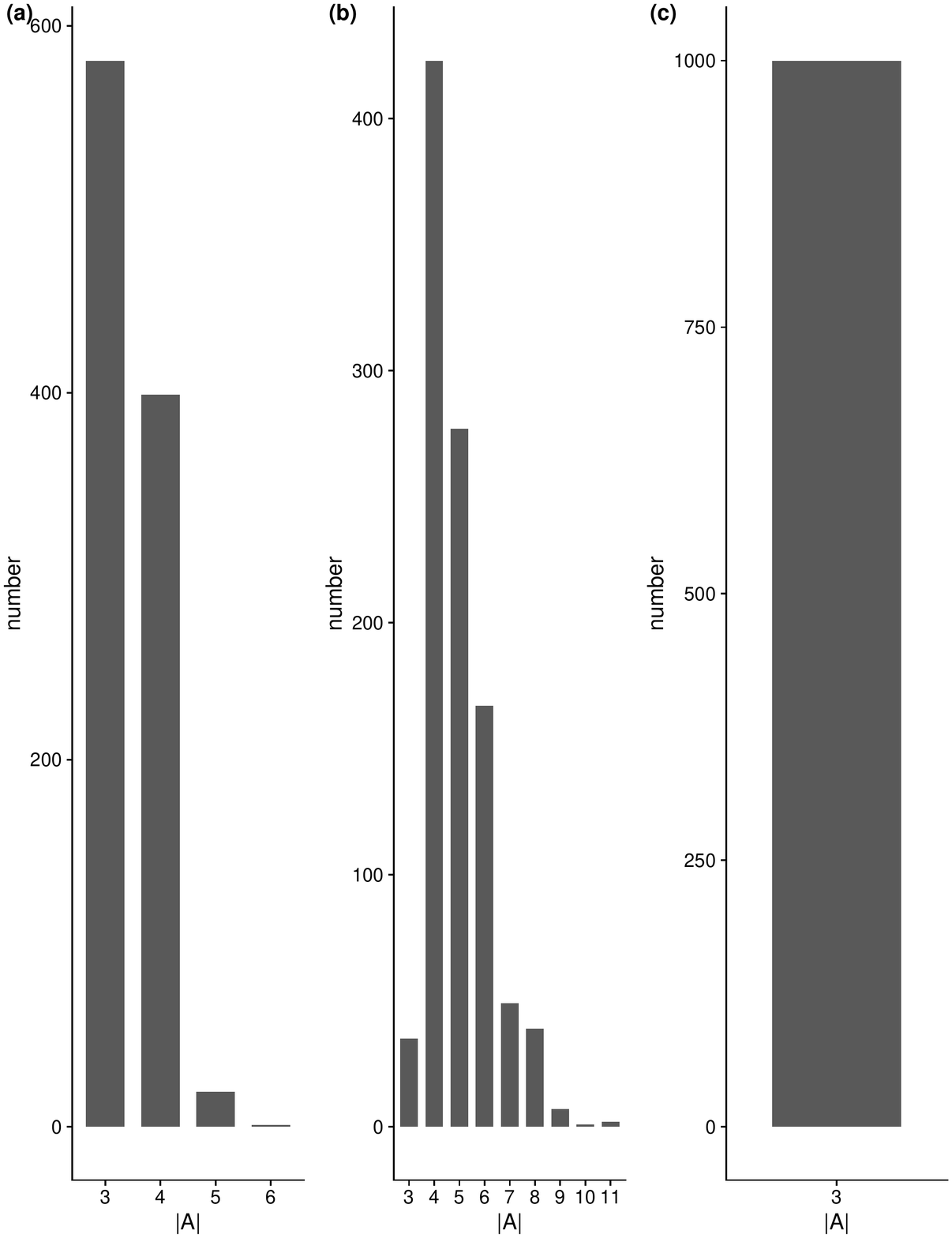}
	\caption{ Histograms for 1000 Monte Carlo replications of $|\widehat{\cal A}_m|$ for adaptive LASSO expectile (a), adaptive LASSO quantile(b) and adaptive LASSO  LS (c) estimations, when the design  $\eX$ is D2,   $m=300$, $p=250$, $|{\cal A}^0|=3$, $\varepsilon \sim{\cal N}(0,1)$.}
	\label{fig_A_N_L2} 
\end{figure}

\begin{figure}[h!]
	\includegraphics[width=15cm,height=5cm,angle=0]{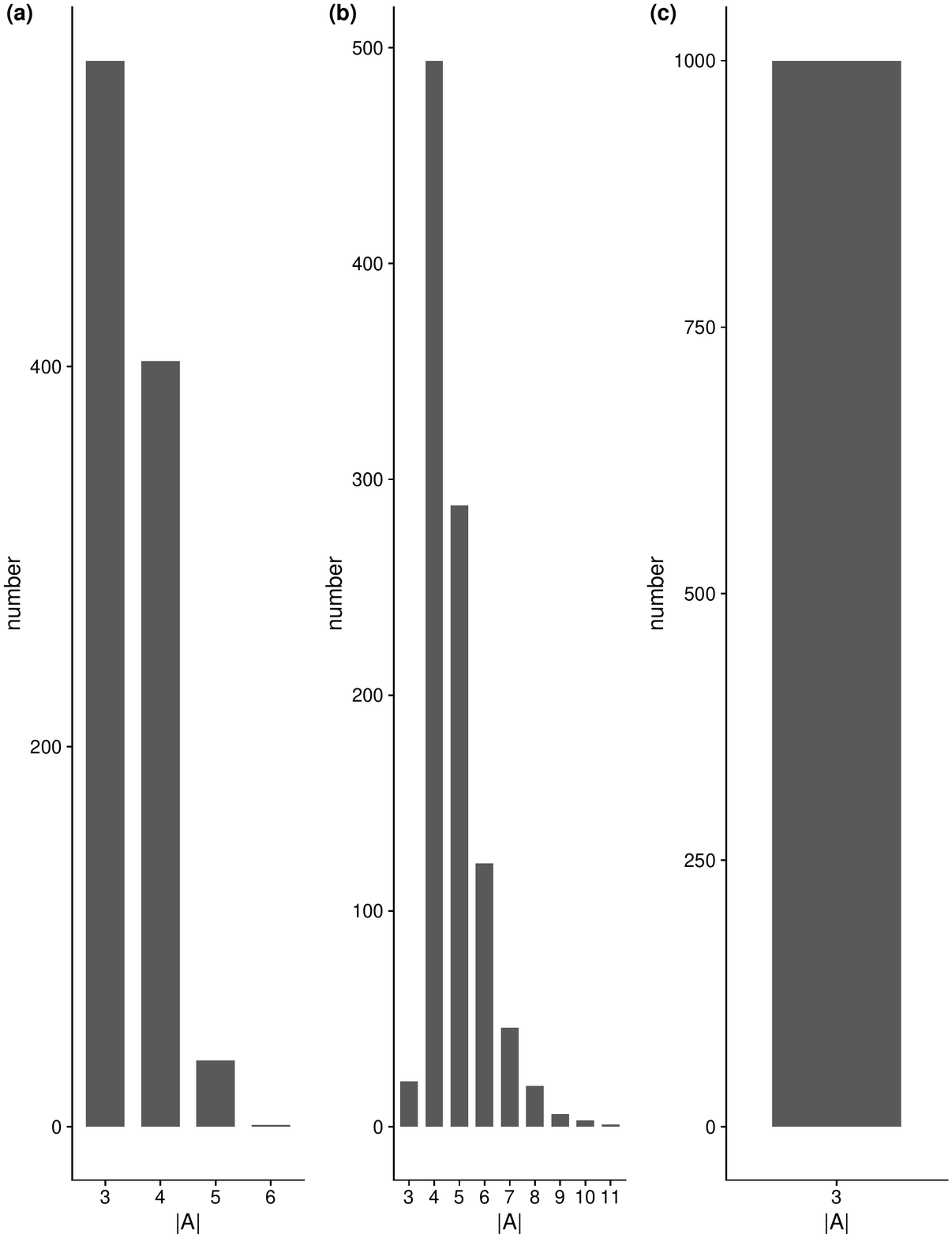}
	\caption{ Histograms for 1000 Monte Carlo replications of $|\widehat{\cal A}_m|$ for adaptive LASSO expectile (a), adaptive LASSO quantile(b) and adaptive LASSO  LS (c) estimations, when the design  $\eX$ is D2,   $m=300$, $p=250$, $|{\cal A}^0|=3$, $\varepsilon \sim{\cal E}\!xp(-1.5)$.}
	\label{fig_A_E_L2} 
\end{figure}

To complete the comparison of the three automatic variable selection methods, in  Figures \ref{fig_A_N_L1} and \ref{fig_A_E_L1}, we represent  $|\widehat{\cal A}^*_m|$ for expectile, quantile and LS methods with adaptive LASSO penalties, for $m=300$, $p=250$, design D1 and 1000 Monte Carlo replications.  The errors are  Normal   in the Figure  \ref{fig_A_N_L1} and of distribution ${\cal E}\!xp(-1.5)$ in Figure \ref{fig_A_E_L1}. Similarly we have the histograms of  Figures  \ref{fig_A_N_L2} and \ref{fig_A_E_L2} when the design is D2. We observe that by adaptive LASSO LS method estimation, we obtain  $|\widehat{\cal A}^*_m|=|{\cal A}^0| =3$, for each   Monte Carlo replication. By the adaptive LASSO expectile method, we identify  the exact number of non-zero coefficients in most cases, while by the adaptive LASSO quantile method, the exact number of three non-zero coefficients is very rarely found, being overestimated, $|\widehat{\cal A}^*_m|$ takes values from 3 to 15. We deduce that the penalized expectile estimation method is superior to that quantile from the point of view of the identification of the true non-zero coefficients when $p$ is very close to $m$.

\subsubsection{Stopping time estimation}
In Table \ref{Ehatk} we give some summary statistics (\textit{min, median, max}) of the stopping times obtained by 2000 Monte Carlo replications for the test statistics constructed on:
 	\begin{itemize}
 		\item adaptive LASSO expectile residuals;
 		\item expectile residuals (only for $p=10$);
 		\item adaptive LASSO quantile residuals;
 		\item modified adaptive LASSO LS residuals.
 	\end{itemize}
 Two possible values for $p$ are considered $p \in \{10, 100\}$, two model errors ${\cal N}(0,1)$, $ {\cal E}\!xp(-1.5)$ and the two designs D1 and D2. We get the same median   of the stopping times by the statistics $\sup_{1 \leqslant k \leqslant T_m}  \Gamma^*(m, k,\gamma)$ and $\sup_{1 \leqslant k \leqslant T_m}  \Gamma(m, k,\gamma)$ when $p=10$. The results obtained by the statistic built on modified adaptive LASSO LS residuals  are in correlation with those of Table \ref{L3_L1} because it does not identify the change for each Monte Carlo replication. Consequently, the median and the maximum of the stopping times are $\infty$ (the change was not detected before observation $T_m=300$). The test statistic built on the adaptive LASSO quantile residuals identifies the change much later than the two test  statistics proposed in this paper. In general, our two statistics detect earlier the change in a model with asymmetric errors compared to the statistics on the adaptive LASSO quantile and modified adaptive LASSO LS residuals.
  \begin{table}
	\caption{\footnotesize Summary statistics for stopping time ($\widehat{k}_m $)  by the test statistics on the adaptive LASSO expectile, expectile,  adaptive LASSO quantile and modified adaptive LASSO LS residuals, for open-end  procedure,  $k^0_m=10$,     $m=T_m=300$ and 2000 Monte Carlo replications.}
	\begin{center}
		{\scriptsize
			\begin{tabular}{|c|c|c|c|ccc|}\hline 
		$	p$ & $\eX$ & $\varepsilon$ &	  estimation &   \multicolumn{3}{c|}{$\widehat{k}_m $} \\ 
				\cline{4-7} 
				  &   & & method &min  & median  & max \\ \hline \hline 
			100 & D2 & ${\cal N}(0,1)$ & ad. LASSO expectile  & 19 & 29 & 43 \\
			 & & &modif ad. LASSO LS & 19 & 25 & 29 \\
			  & & &  ad. LASSO quantile & 48 & 77 & 147 \\
		  \cline{3-7}
			& &${\cal E}\!xp(-1.5)$ & ad. LASSO expectile  & 22 & 24 & 32 \\
			& & &modif ad. LASSO LS & 16 & 22 & 23 \\
			& &  & ad. LASSO quantile & 100 & 159 & 210 \\
			\cline{2-7}
			  & D1 & ${\cal N}(0,1)$ & ad. LASSO expectile  & 26 & 26 & 27 \\
			& & &modif ad. LASSO LS & 74 & $\infty$ & $\infty$ \\
			& & &  ad. LASSO quantile & 67 & 96 & 117 \\
			\cline{3-7}
			& &${\cal E}\!xp(-1.5)$ & ad. LASSO expectile  & 24 & 27 & 28 \\
			& & &modif ad. LASSO LS &  52 & $\infty$ & $\infty$ \\
			& &  & ad. LASSO quantile & 52 & 74 & 103 \\
			\cline{1-7}
				10 & D2 & ${\cal N}(0,1)$ & ad. LASSO expectile  & 12 & 12 & 20 \\
				& & &  expectile & 12 & 12 & 20 \\
			& & &modif ad. LASSO LS & 15 & 18 & 20 \\
			& & &  ad. LASSO quantile & 63 & 98 & 166 \\
			\cline{3-7}
			& &${\cal E}\!xp(-1.5)$ & ad. LASSO expectile  & 24 & 24 & 41 \\
				& & &  expectile & 12 & 24 & 42 \\
			& & &modif ad. LASSO LS & 15 & 17 & 21 \\
			& &  & ad. LASSO quantile & 141 & 193 & 209 \\
			\cline{2-7}
			& D1 & ${\cal N}(0,1)$ & ad. LASSO expectile  & 16 & 24 & 31 \\
				& & &  expectile & 16 & 24 & 31 \\
			& & &modif ad. LASSO LS & 18 & 75 & 84 \\
			& & &  ad. LASSO quantile & 77 & 96 & 128 \\
			\cline{3-7}
			& &${\cal E}\!xp(-1.5)$ & ad. LASSO expectile  & 21 & 31 & 34 \\
				& & &  expectile & 22 & 29 & 34 \\
			& & &modif ad. LASSO LS &  21 & 78 & 95 \\
			& &  & ad. LASSO quantile & 43 & 69 & 89 \\ \hline
			
			\end{tabular} 
		}
	\end{center}
	\label{Ehatk} 
\end{table}

\subsubsection{Conclusions of the simulations} 
The best performing values of $\gamma$ that give best results for the test statistics   $\sup_{1 \leqslant k \leqslant T_m} \Gamma(m, k,\gamma)$ and $\sup_{1 \leqslant k \leqslant T_m} \Gamma^*(m, k,\gamma)$ are of $0$ to $0.15$, the value of the empirical sizes $\widehat{\alpha}$ increasing with $\gamma$. If there are no zero model coefficients then, the test statistic  $\sup_{1 \leqslant k \leqslant T_m} \Gamma(m, k,\gamma)$ built on the expectile residuals make fewer false point-change detections than $\sup_{1 \leqslant k \leqslant T_m} \Gamma^*(m, k,\gamma)$. Conversely, if the model has irrelevant variables then the statistic $\sup_{1 \leqslant k \leqslant T_m} \Gamma^*(m, k,\gamma)$ gives better results, in terms of  $\widehat{\alpha}$, than $\sup_{1 \leqslant k \leqslant T_m} \Gamma(m, k,\gamma)$, especially when $p$ is large and close to $m$.  Under hypothesis $H_0$, the obtained values for the empirical size $\widehat{\alpha}$ by the open-end procedure are lower than those obtained by the closed-end procedure. For a model with a small or large number of explanatory variables, some of which are irrelevant, the  test statistics constructed on the   adaptive LASSO expectile residuals, adaptive LASSO quantile residuals and modified adaptive LASSO LS residuals, when the hypothesis $H_0$ is true, that is no change after historical data,  give similar results for symmetrical or asymmetrical model errors. Indeed, by the three methods we obtained $\widehat{\alpha}  < 0.05$ when $\gamma \in \{0, 0.15 \}$.  In exchange, if the model changes after the historical data, the test statistics  on the adaptive LASSO quantile residuals and on the modified adaptive LASSO LS  residuals do not detect the change every time, especially when the  model errors have an asymmetric distribution. \\
On the other hand, the adaptive LASSO expectile estimation method is superior to that of adaptive LASSO quantile estimation method from the point of view of the identification of true non zero coefficients when the number of explanatory variables is very close to the number of historical observations.\\
The  detection delays of the change under hypothesis $H_1$ in a model with asymmetric errors is shorter for our two test statistics than those given by the two comparison test statistics.\\
All of these, show that our proposed test statistic gives very good results for real-time detection of a change-point and above all it is superior to the statistics constructed on the residuals of the adaptive LASSO quantile and modified adaptive LASSO LS  methods, especially when the model error  distribution   is asymmetrical.
\subsection{Applications on real data}
\label{subsect_application}
This subsection presents two applications on the real data of the test statistics $\sup_{1\leqslant k \leqslant T_m} \Gamma(m,k,\gamma) $ and $\sup_{1 \leq k \leq T_m}  \Gamma^*(m,k,\gamma)$ by the open-end procedure.
\subsubsection{Aquatic toxicity towards the fishes}
\begin{figure}[h!]
	\includegraphics[width=16cm,height=6cm,angle=0]{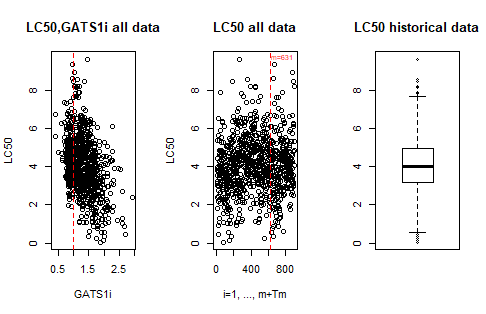}
	\caption{The aquatic acute  toxicity concentration LC50 .}
	\label{fish} 
\end{figure}

In the first example, the aquatic acute  toxicity towards the fish Pimephales promelas is studied. The   data proposed by \cite{Cassotti.15}   can be downloaded from the \textit{Machine Learning Repository} site \href{https://archive.ics.uci.edu/ml/datasets/QSAR+fish+toxicity\#}{https://archive.ics.uci.edu/ml/datasets/QSAR+fish+toxicity\#}. The response variable $Y$ is the  aquatic acute  toxicity concentration, denoted by LC50. There are six explanatory variables on the molecular descriptors: MLOGP (molecular properties), CIC0 (information indices), GATS1i (2D autocorrelations), NdssC (atom-type counts), NdsCH ((atom-type counts), SM1\_Dz (2D matrix-based descriptors). The number of observations in the database is 908. In the left sub-figure of Figure \ref{fish} we represent LC50 in respect to GATS1i on the all 908 observations. Taking into account this sub-figure, we reorder the database with respect to the decreasing values of GATS1i and we take as historical observations those for which GATS1i>1. The number of historical observations is $m = 631$ and those from $T_m=277$ (see the middle sub-figure of Figure \ref{fish}, where the dotted line is for $m=631$). In the right sub-figure of Figure \ref{fish} we have the boxplot of LC50 on the historical observations. We observe that the values of LC50 are asymmetrical around the empirical average, the estimated value of the expectile index $\tau$ being 0.469. For $i=1, \cdots , 908$, the explanatory variable vector is $\eX_i=(MLOGP_i,CIC0_i,GATS1i_i,NdssC_i,NdsCH_i,SM1\_Dz_i)$. We obtain that the expectile estimation of the model coefficients on the historical observations is $\widehat{\eb}_m=(0.34,  1.32, -0.85,  0.43,  0.02,$ $  0.43)^\top$, while the adaptive LASSO expectile estimation is $\widehat{\eb}^*_m=(0.05,  1.05, -0.69,  0.18, 0,  0.44)^\top$. Since the coefficient of NdssC in $\widehat{\eb}^*_m$ is zero we deduce that this number of atoms does not influence the concentration of LC50 and then $\widehat{\cal A}^*_m=\{1, 2, 3, 4, 6\}$. By the two test statistics $\sup_{1\leqslant k \leqslant T_m} \Gamma(m,k,\gamma) $ and $\sup_{1 \leq k \leq T_m}  \Gamma^*(m,k,\gamma)$  we get that there is a change in the model after the historical data. The stopping times are estimated as 30 and 22 by expectile and adaptive LASSO expectile test statistics, respectively. This means that by the test statistic based on the expectile residuals, the model will change as soon as the value of GATS1i is smaller than 0.963 and by the test statistic based on the adaptive LASSO expectile residuals, the model changes as soon as $GATS1i \leq 0.975$.
\subsubsection{NO2 pollution data}
\begin{figure}[h!]
  	\includegraphics[width=16cm,height=6cm,angle=0]{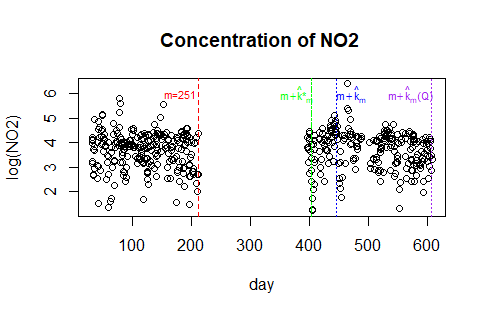}
  	\caption{Observations for the logarithm of the NO2.}
  	\label{NO2} 
  \end{figure}

In this second example we take the one presented in \cite{Ciuperca-18} where a test statistic built on the adaptive LASSO quantile estimator was used to detect a change in a linear model of NO2 pollution with respect to six explanatory variables:  logarithm of the number of cars per hour,    temperature 2 m above ground,    wind speed,   temperature difference between 25 and 2 m above ground,  wind direction,  hour of day when the data were measured. Note that there may be days with several measures and days without measures. Historical data measured between  November 1, 2001 and April 30, 2002, contains 251 observations for 212 days. From May 1, 2002 there are no data until October 31 (inclusive), 2002 and therefore there is a gap in Figure \ref{NO2}.   From November 1, 2002 other $T_m=249$ were made for 211 days, until August 2003. The all values of NO2 are represented in Figure \ref{NO2} where with the red dotted vertical line we represented the observation $m=251$. The data come from the Dept. of Statistics, Carnegie Mellon University and can be uploaded to  \href{http://lib.stat.cmu.edu/datasets/NO2.dat}{http://lib.stat.cmu.edu/datasets/NO2.dat}. By the adaptive LASSO quantile estimation method, \cite{Ciuperca-18} got that the number of cars per hour and  the temperature difference between 25 and 2 m  are relevant variables for NO2. The associated test statistic detects a change to observation 248 after the historical data. For the estimated expectile  index $\widehat \tau$ equal to 0.62, we obtain that only adaptive LASSO estimation of the coefficient of the number of cars is different from zero. The test statistics $\sup_{1\leqslant k \leqslant T_m} \Gamma(m,k,\gamma) $ detects a change to observation 70 after the historical data and $\sup_{1 \leq k \leq T_m}  \Gamma^*(m,k,\gamma)$ to observation 14. In Figure \ref{NO2} we have drawn with dotted lines in green and blue the observations where the change is detected by  $\sup_{1 \leq k \leq T_m}  \Gamma^*(m,k,\gamma)$ and  $\sup_{1 \leq k \leq T_m}  \Gamma(m,k,\gamma)$, respectively. Then, the two test statistics proposed in the present paper  detect earlier the change in the model compared to the change found by the statistic based on the adaptive LASSO quantile estimator proposed by \cite{Ciuperca-18}, drawn with a vertical purple dotted line in Figure \ref{NO2}. 
The results are summarized in Table \ref{T_NO2}. In Table \ref{T_NO2} and Figure \ref{NO2}, $\widehat{k}^*_m$ is the stopping time of relation (\ref{km*}) obtained by $\sup_{1 \leq k \leq T_m}  \Gamma^*(m,k,\gamma)$, ${\widehat{k}_m } $ of (\ref{km}) by $\sup_{1 \leq k \leq T_m}  \Gamma(m,k,\gamma)$ and $\widehat{k}_m(Q)$ by test statistic built on the adaptive LASSO quantile residuals of \cite{Ciuperca-18}.
 \begin{table}
	\caption{\footnotesize Last historical day and days of changes by three statistics.}
	\begin{center}
		{\scriptsize
			\begin{tabular}{|c|c|c|c|}\hline 
				 $Day[m]$ & $m+{\widehat{k}^*_m } $ & $m+{\widehat{k}_m } $ & $m+\widehat{k}_m(Q)$\\ 
				\hline \hline 
				April 30, 2002 & November 8, 2002 & December 20, 2002 & July 30, 2003 \\ \hline
			\end{tabular} 
		}
	\end{center}
	\label{T_NO2} 
\end{table}

\section{Proofs}
\label{sect_proof}
In this section we present the proofs of the results stated in Sections \ref{sect_models} and \ref{sect_results}.\\

{\bf Proof of Proposition  \ref{Propositon 1}.} 
The expectile estimator $\widehat{\eb}_m$ is the solution of the system of equations: $\frac{\partial}{\partial \eb}\sum^m_{i=1} \rho_\tau(Y_i-\eX_i^\top \eb )=\textbf{0}_p$. Then:
\begin{equation}
\label{eq2}
\sum^m_{i=1} g_\tau(\varepsilon_i-\eX_i^\top (\widehat{\eb}_m - \eb^0)) \eX_i=\textbf{0}_p.
\end{equation}
On the other hand, by elementary calculations, for $t \rightarrow 0$, we have:
\begin{equation}
\label{eq3}
g_\tau(\varepsilon-t)=g_\tau(\varepsilon) - h_\tau(\varepsilon) t+o_\PP(t).
\end{equation}
Taking into account assumption   (A1) and since $ \widehat{\eb}_m \overset{{\PP}} {\underset{m \rightarrow \infty}{\longrightarrow}} \eb^0$, we have for $t=\eX_i^\top (\widehat{\eb}_m - \eb^0)$, $i=1 , \cdots , m$:
\[
g_\tau \big(\varepsilon_i-\eX_i^\top (\widehat{\eb}_m - \eb^0) \big)=g_\tau(\varepsilon_i)-h_\tau(\varepsilon_i)\eX_i^\top (\widehat{\eb}_m - \eb^0)+o_\PP \big( \eX_i^\top (\widehat{\eb}_m - \eb^0)\big). 
\]
Then, we can write relation (\ref{eq2}):
\begin{equation}
\label{eq4}
\textbf{0}_p = \sum^m_{i=1}g_\tau(\varepsilon_i)\eX_i - \sum^m_{i=1} h_\tau(\varepsilon_i)\eX_i\eX_i^\top (\widehat{\eb}_m - \eb^0)+\sum^m_{i=1} \eX_i o_\PP \big(\eX_i^\top (\widehat{\eb}_m - \eb^0)\big).
\end{equation}
By assumptions (A2) and (A3) and since $\mu_h >0$, we have by the law of large numbers (LLN):
\begin{equation}
\label{eq5}
\frac{1}{m}   \sum^m_{i=1} h_\tau(\varepsilon_i)\eX_i\eX_i^\top  \overset{{\PP}} {\underset{m \rightarrow \infty}{\longrightarrow}} \mu_h \eO.
\end{equation}
On the other hand, since $\eE[g_\tau(\varepsilon)]=0$, together with assumption (A1), by the Bienaymé-Tchebychev inequality, we have:
\begin{equation}
\label{eq6}
\big\|\sum^m_{i=1}g_\tau(\varepsilon_i)\eX_i\big\|_2=O_\PP(m^{1/2}).
\end{equation}
Combining relations (\ref{eq4}), (\ref{eq5}), (\ref{eq6}) and assumptions (A1), (A3), we obtain:
\begin{equation*} 
\begin{split}
\label{eq7}
\widehat{\eb}_m - \eb^0 &= \bigg(\frac{1}{m}    \sum^m_{i=1} h_\tau(\varepsilon_i)\eX_i\eX_i^\top\bigg)^{-1}\bigg( \frac{1}{m} \sum^m_{i=1}g_\tau(\varepsilon_i)\eX_i + o_\PP(m^{-1/2})\bigg)\\
& =\mu^{-1}_h  \eO^{-1} \frac{1}{m}  \sum^m_{i=1}g_\tau(\varepsilon_i)\eX_i  +o_\PP(m^{-1/2}).
\end{split}
\end{equation*}
The proposition is proved.
\hspace*{\fill}$\blacksquare$  \\

Recall the Hoeffding's inequality which will be used in the proof of  Lemma  \ref{Lemma 4.2 (SQA)}.\\

\textbf{Hoeffding's inequality} (\cite{Hoeffding.63}): \textit{If $(Z_i)_{1 \leqslant i \leqslant n}$  are independent random variables,  $(a_i)_{1 \leqslant i \leqslant n}$ and $(b_i)_{1 \leqslant i \leqslant n}$ two real sequences with $a_i < b_i$ and  $\PP[a_i \leq Z_i \leq b_i]=1$, for any $i =1, \cdots , n$,  then for all $t>0$:
	\[
	\PP \big[\big|\sum^n_{i=1} \big(Z_i -\eE[Z_i]\big) \big| \geq t \big] \leq 2 \exp \bigg(- \frac{2t^2}{\sum^n_{i=1}(b_i-a_i)^2}\bigg).
	\]}

{\bf Proof of Lemma \ref{Lemma 4.2 (SQA)}.}
For all $x \in \R$, we can write  the function $g_\tau(x)$ as: $g_\tau(x)=2 \tau x +2(1-\tau)x \e1_{x <0}$.Then,  for $\eu \in \R^p$, with $\| \eu \|_2 \leq C_1$ and $C_1$ a positive constant, we have for $i=m+1, \cdots , m+k$,
\[
\frac{1}{2} \big(g_\tau \big(\varepsilon_i -m^{-1/2} \eX_i^\top  \eu \big)- g_\tau(\varepsilon_i)\big)= -m^{-1/2} \eX_i^\top  \eu+(1-\tau)\bigg( (\varepsilon_i - m^{-1/2} \eX_i^\top  \eu ) \e1_{\varepsilon_i \leq m^{-1/2} \eX_i^\top  \eu}- \varepsilon_i \e1_{\varepsilon_i \leq 0} \bigg).
\]
Hence, for $j \in \{1, \cdots , p \}$, the $j$-th component of the random process $\eR_i(\eu)$ is:
\[
R_{ij}(\eu)= 2 X_{ij} \bigg[-m^{-1/2} \eX_i^\top  \eu +(1-\tau) \bigg(-m^{-1/2} \eX_i^\top  \eu \e1_{\varepsilon_i \leq m^{-1/2} \eX_i^\top  \eu} +\varepsilon_i (\e1_{\varepsilon_i \leq m^{-1/2} \eX_i^\top  \eu}-\e1_{\varepsilon_i \leq 0}).
\bigg)\bigg]
\]
In order to study $R_{ij}(\eu)$, let us consider  the random variable: $W_i \equiv \varepsilon_i (\e1_{\varepsilon_i \leq m^{-1/2} \eX_i^\top  \eu}-\e1_{\varepsilon_i \leq 0}) $. \\
Using the identity that, for all $a, b \in \R$:  $|\e1_{\varepsilon \leq a} - \e1_{\varepsilon \leq b}| =\e1_{\min(a,b) \leq \varepsilon \leq \max(a,b)}$, then we can write: $ |W_i|=|\varepsilon_i| \e1_{\min(0,m^{-1/2} \eX_i^\top  \eu) \leq \varepsilon_i \leq \max(0,m^{-1/2} \eX_i^\top  \eu)} $. Thus, we have: $\PP[|W_i| \leq m^{-1/2}| \eX_i^\top  \eu |]=1$. Furthermore, using assumption (A1) we have that the component $R_{ij}(\eu)$ is bounded with probability 1, by $C_3 m^{-1/2}$, with $C_3 >0$ a constant. Let's take in Hoeffding inequality $Z_i=R_{ij}(\eu)$, for $i=m+1, \cdots , m+k$ and $b_i=-a_i=C_3 m^{-1/2}$. Then, for all $t>0$ we have:
\begin{equation}
\label{eq8}
\PP \big[\big|\sum^{m+k}_{i=m+1} (R_{ij}(\eu)-\eE[R_{ij}(\eu)]) \geq t\big] \leq 2 \exp \bigg(- \frac{ 2 t^2}{4 C_3^2 k m^{-1}}\bigg).
\end{equation}
We take $t= 2^{1/2} C_2 C_3  (k/m)^{1/2}  (\log k)^{1/2}$, for any positive constant $C_2$. Then,  relation (\ref{eq8}) becomes, for all $j \in \{1, \cdots , p\}$:
\begin{equation}
\label{eq9}
\PP \bigg[\big|\sum^{m+k}_{i=m+1} \bigg(R_{ij}(\eu)-\eE[R_{ij}(\eu)]\bigg) \geq  \sqrt{2} C_2 C_3 \bigg(\frac{k}{m} \log k\bigg)^{1/2} \bigg] \leq 2 \exp(-C_2^2 \log k)=2 k^{-C_2^2}.
\end{equation}
Thus, relation (\ref{eq9}) implies that for all $C_1 >0$ such that $\| \eu \|_2 \leq C_1$, for all $C_2 >0$, there exists a constant $C_3 >0$ such that:
\begin{eqnarray*}
	\PP \bigg[\big\|\rr_{m,k}(\eu) - \eE[\rr_{m,k}(\eu)] \big\|_1 \geq   2^{1/2} C_2 C_3 p m^{-1/2} k^{1/2}   (\log{k})^{1/2} \bigg] \qquad \qquad \qquad \qquad \qquad  \qquad\\
	\leq \PP \bigg[\max_{1 \leqslant j \leqslant p} \bigg|\sum^{m+k}_{i=m+1} \bigg( R_{ij}(\eu) - \eE[R_{ij}(\eu)]\bigg) \bigg| \geq  2^{1/2} C_2 C_3 m^{-1/2} k^{1/2} (\log{k})^{1/2} \bigg] \leq 2 k^{-C_2^2},
\end{eqnarray*}
and Lemma follows.
\hspace*{\fill}$\blacksquare$  \\

{\bf Proof of Theorem \ref{Theorem 2.1(SQA)}.}
We first show that the supremum of $\Gamma(m,k,\gamma)$  cannot be reached for $k$ small. Effectively, if $k$ is small then  $z(m,k,\gamma) \simeq m^{1/2} \big(1+  m^{-1}\big) \big( 1+m\big)^{-\gamma} {\underset{m \rightarrow \infty}{\longrightarrow}}  \infty$. On the other hand,  since $ \widehat{\eb}_m \overset{{\PP}} {\underset{m \rightarrow \infty}{\longrightarrow}} \eb^0$, taking into account  assumptions (A1) and (A3), the fact that $g_\tau(.)$ is a continuous Borel function, then we have that: $g_\tau(Y_i -\eX_i^\top\widehat{\eb}_m )  \overset{{\PP}} {\underset{m \rightarrow \infty}{\longrightarrow}} 0 $. Thus, taking into account  assumption (A1), we have: $ g_\tau(Y_i -\eX_i^\top\widehat{\eb}_m ) \eX_i \overset{{\PP}} {\underset{m \rightarrow \infty}{\longrightarrow}} \textbf{0}_p $ and therefore $\|\ \eJ_m^{1/2} \sum^{m+k}_{i=m+1} g_\tau(Y_i - \eX_i^\top\widehat{\eb}_m ) \eX_i  \|_\infty  \overset{{\PP}} {\underset{m \rightarrow \infty}{\longrightarrow}} 0$. Hence, for $k$ small we have,  $\Gamma(m,k,\gamma) \overset{{\PP}} {\underset{m \rightarrow \infty}{\longrightarrow}} 0$.\\
We therefore consider that $k \rightarrow\infty$ and we study first $  \frac{m^{-1/2} k^{1/2} (\log k)^{1/2}}{z(m,k,\gamma)}$ in  two cases.
\begin{description}
	\item \textit{a) if $   k \leq m$}, then, taking into account that $0 \leq \gamma < 1/2$,  we have:
	\begin{align*} 
	\frac{m^{-1/2} k^{1/2} (\log k)^{1/2}}{z(m,k,\gamma)}&=\frac{k^{1/2}}{m \big( 1+{k}/{m}\big)} (\log k)^{1/2} \frac{\big(1+{k}/{m}\big)^\gamma}{\big({k}/{m}\big)^\gamma} \leq 2^\gamma \frac{k^{1/2 - \gamma} (\log k)^{1/2} m^{\gamma -1}}{1+{k}/{m}}\\
	&  \leq 2^\gamma m^{1/2 - \gamma} m^{\gamma -1} (\log m)^{1/2} = 2^\gamma m^{-1/2} (\log m)^{1/2} {\underset{m \rightarrow \infty}{\longrightarrow}} 0.  
	\end{align*}
	\item \textit{b) If $m < k \leq \infty $}, then $(\log k)^{1/2} < k^{1/4}$, from where, 	with the notation $\zeta=k/m >1$, we get:
	\begin{align*}
	\frac{m^{-1/2} k^{1/2} (\log k)^{1/2}}{z(m,k,\gamma)}& \leq \left(\frac{k}{m} \right)^{1/2} \frac{k^{1/4} m^{-1/2}}{\big(1+\frac{k}{m}\big) \big(\frac{k/m}{1+k/m}\big)^\gamma}= \frac{\zeta^{3/4 - \gamma}}{(1+\zeta)^{1-\gamma}}m^{-1/4}\\
	 & = m^{-1/4}\bigg(\frac{\zeta}{1+\zeta}\bigg)^{3/4- \gamma} (1+\zeta)^{-1/4}.
		\end{align*}
	But $\zeta(1+\zeta)^{-1} <1$ and $\gamma <3/4$. Moreover $(1+\zeta)^{-1} <2^{-1}$, from where $(1+\zeta)^{-1/4} < 2^{-1/4}$. These last two relations imply $ \frac{m^{-1/2} k^{1/2} (\log k)^{1/2}}{z(m,k,\gamma)} \leq 2^{-1/4} m^{-1/4}=o(1)$. 
\end{description}
Thus, we showed in cases \textit{a)} and \textit{b)}  that $ \frac{m^{-1/2} k^{1/2}(\log k)^{1/2}}{z(m,k,\gamma)}=o(1)$. On the other hand, by Lemma \ref{Lemma 4.2 (SQA)} we have, for $\eu \in \R^p$, $\| \eu \|_2 \leq C_1< \infty$, that:
\begin{equation}
\label{eq12}
\rr_{m,k}(\eu)=\eE[\rr_{m,k}(\eu)]+O_\PP \big(m^{-1/2} k^{1/2} (\log k)^{1/2}\big).
\end{equation}
For the expectation in the right-hand side of (\ref{eq12}), since hypothesis $H_0$ is true, using relation (\ref{eq3}), together with the supposition that $\eE[g_\tau(\varepsilon_i)]=0$ of (A3) and assumption (A1), we have:
\[ 
\eE[\rr_{m,k}(\eu)]=\sum^{m+k}_{i=m+1} \eX_i \eE \big[- h_\tau(\varepsilon_i) \eX_i^\top m^{-1/2} \eu +o_\PP(m^{-1/2})\big]=- m^{-1/2} \mu_h \sum_{m+k}^{i=m+1} \eX_i \eX_i^\top \eu + o(m^{-1/2} k).
\]
We replace this last relation in (\ref{eq12}) and we obtain:
\begin{equation}
\label{eq13}
\rr_{m,k}(\eu)=- m^{-1/2} \mu_h \sum_{m+k}^{i=m+1} \eX_i \eX_i^\top \eu + o(m^{-1/2} k)+O_\PP \big(m^{-1/2} k^{1/2} (\log k)^{1/2}\big).
\end{equation}
On the other hand, by Proposition \ref{Propositon 1}, we have:
\[
m^{1/2}(\widehat{\eb}_m - \eb^0)=m^{-1/2} \mu^{-1}_h \eO^{-1}\sum^{m}_{i=1}g_\tau(\varepsilon_i) \eX_i +o_\PP(1).
\] 
We take then   $\eu=m^{1/2}(\widehat{\eb}_m - \eb^0)$  and relation (\ref{eq13}) becomes:
\begin{equation}
 \begin{split}
\label{eq14}
\rr_{m,k}(m^{1/2}(\widehat{\eb}_m - \eb^0))=- m^{-1} \eO^{-1}  \big( \sum^{m+k}_{i=m+1} \eX_i \eX_i^\top\big) \big(\sum^{m}_{j=1} g_\tau(\varepsilon_j)\eX_j \big) \\
+ o(m^{-1/2} k)+O_\PP \big(m^{-1/2} k^{1/2} (\log k)^{1/2}\big).
\end{split}
\end{equation}
But, by assumption (A2), since $k$ is large, we have: $\sum^{m+k}_{i=m+1} \eX_i \eX_i^\top =k \eO (1+o(1))$. Thus, relation (\ref{eq14}) becomes:
\begin{equation*}
\label{eq15} 
\rr_{m,k}(m^{1/2}(\widehat{\eb}_m - \eb^0))=-k  m^{-1} \sum^{m}_{j=1} g_\tau(\varepsilon_j)\eX_j  + o(m^{-1/2} k)+O_\PP \big(m^{-1/2} k^{1/2} (\log k)^{1/2}\big).
\end{equation*}
This relation implies, taking into account the definition of $\rr_{m,k}$:
\begin{equation}
\begin{split}
\label{eq16}
\sum^{m+k}_{i=m+1}  g_\tau(\widehat{\varepsilon}_i )\eX_i = \sum^{m+k}_{i=m+1}  g_\tau( \varepsilon_i)\eX_i- k  m^{-1} \sum^{m}_{j=1} g_\tau(\varepsilon_j)\eX_j \\
  + o(m^{-1/2} k)+O_\PP \big(m^{-1/2} k^{1/2} (\log k)^{1/2}\big).
\end{split}
\end{equation} 
According to the K-M-T approximation,  there exists two independent $p$-dimensional  Wiener processes $\big\{\textbf{W}_{1,m}(t), t \in [0,\infty)\big\}$ and $\big\{\textbf{W}_{2,m}(t), t \in [0, \infty ) \big\}$ such that, for $\nu >2$ and $m \rightarrow \infty$ (see \cite{Komlos.Major.Tusnady.75} and \cite{Komlos.Major.Tusnady.76}):
\begin{equation}
\label{eq17}
\sup_{1 \leq k < \infty} k^{-1/\nu} \big\| \eJ_m^{-1/2} \sum^{m+k}_{i=m+1}  g_\tau( \varepsilon_i) \eX_i- \textbf{W}_{1,m}(k)\big\|_\infty =O_\PP(1)
\end{equation}
and
\begin{equation}
\label{eq18}
\big\| \eJ_m^{-1/2}\sum^{m}_{j=1} g_\tau(\varepsilon_j)\eX_j   -\textbf{W}_{2,m}(m) \big\|_\infty =o_\PP(m^{1/ \nu}).
\end{equation}
Then, taking into account relations (\ref{eq16}), (\ref{eq17}) and (\ref{eq18}), we obtain:
\begin{equation}
\label{to}
\sup_{1 \leq k < \infty} \frac{\bigg\|  \eJ_m^{-1/2} \sum^{m+k}_{i=m+1}  g_\tau( \varepsilon_i)\eX_i -\frac{k}{m} \eJ_m^{-1/2}\sum^{m}_{j=1} g_\tau(\varepsilon_j)\eX_j - \bigg( \textbf{W}_{1,m}(k)- \frac{k}{m} \textbf{W}_{2,m}(m)\bigg)\bigg\|_\infty}{z(m,k,\gamma)} =o_\PP(1).
\end{equation}
On the other hand,  as in the proof of Theorem 2.1 of \cite{Horvath.Huskova.Kokoszka.Steinebach.04} we have that:
\begin{equation}
\label{eq19}
\sup_{1 \leqslant k \leqslant T_m} \frac{\big\|\textbf{W}_{1,m}(k)- \frac{k}{m} \textbf{W}_{2,m}(m) \big\|_\infty }{z(m,k,\gamma)} \overset{\cal L} {\underset{m \rightarrow \infty}{\longrightarrow}}  \left\{
\begin{array}{lll}
\displaystyle{\sup_{0 < t < 1} \frac{\| \textbf{W}_p(t) \|_\infty}{t^\gamma}},& & \textrm{for open-end procedure}\\
\displaystyle{\sup_{0 < t < T/(1+T)} \frac{\| \textbf{W}_p(t) \|_\infty}{t^\gamma}},& & \textrm{for closed-end procedure,}
\end{array}
\right.
\end{equation}
with $\big\{ \textbf{W}_p(t), t\in[0, \infty) \big\}$ a Wiener process of dimension $p$. 
It remains to study the two remainders in relation (\ref{eq16}). By Lemma 2 of \cite{Zou-Wang-Tang.15} we have:
\begin{equation}
\label{eq21}
\frac{o(k m^{-1/2})}{z(m,k,\gamma)} {\underset{m \rightarrow \infty}{\longrightarrow}} 0.
\end{equation}
For the second remainder, we have:
\begin{equation}
\label{eq20}
\frac{O_\PP(k^{1/2} m^{-1/2} (\log k)^{1/2})}{z(m,k,\gamma)}=O_\PP \bigg(\frac{(k/m)^{1/2} (\log k)^{1/2}}{z(m,k,\gamma)} \bigg) .
\end{equation}
\begin{description}
	\item   \underline{If $1 \leq k \leq m$}, then $1/m \leq k/m \leq 1$ and also:
	\begin{description}
		\item \textit{a)}  $1+k/m >1$,
		\item \textit{b)} $k/(k+m)=(k/m)(1+k/m) \leq 1/(1+k/m) <1$,
		\item  \textit{c)} $ m^{-1/2} ( \log k)^{1/2} \leq m^{-1/2} (\log m)^{1/2} $.
	\end{description}
	From \textit{a), b), c)} we obtain that in the case  $1 \leq k \leq m$,  relation (\ref{eq20}) is $o_\PP(1)$  for $m \rightarrow \infty$. 
	\item \underline{If $m < k < \infty$}, using Lemma 2 of  \cite{Zou-Wang-Tang.15}, we have:
	\[
	\frac{(k/m)^{1/2} (\log k)^{1/2}}{m^{1/2}\big(1+ \frac{k}{m}\big) \big(\frac{k}{k+m}\big)^\gamma}= \frac{k^{1/2} m^{-1/4}(\log k)^{1/2}}{z(m,k,\gamma)} m^{-1/4}=o(1) m^{-1/4}=o(1).
	\]
\end{description}
In conclusion we obtain that relation (\ref{eq20}) is $o_\PP(1)$. Combining this with relations (\ref{eq21}), (\ref{eq19}), (\ref{to}) and (\ref{eq16}), the theorem follows.
\hspace*{\fill}$\blacksquare$  \\

{\bf Proof of Theorem \ref{Theorem 2.2(SQA)}.}
We prove the theorem for the  open-end procedure.  The proof for the  closed-end procedure  is similar. Without reducing the generality, we suppose that  $k^0_m \leq m^s/2$, with $s>1$ and we will show that there is an observation $\widetilde{k}_m$ for which the test statistic diverges. We consider  $\widetilde{k}_m= k^0_m +m^s$ and we study the following random vector:
\begin{equation}
\label{eq22}
\eJ^{-1/2}_m \sum^{m+\widetilde{k}_m}_{i=m+1}  g_\tau (Y_i - \eX_i^\top \widehat{\eb}_m)\eX_i.
\end{equation}
Similarly to the proof of Theorem \ref{Theorem 2.1(SQA)}, we have that there exists a  constant $C>0$ such that, 
\[
\frac{\big\| \eJ^{-1/2}_m \sum^{m+ {k}^0_m}_{i=m+1}  g_\tau (Y_i - \eX_i^\top \widehat{\eb}_m)\eX_i\big\|_\infty }{ z(m, k^0_m, \gamma)} \leq C,
\]
with probability converging to  1 as $m \rightarrow \infty $. \\
Since the function $(1+x)(x/(1+x))^\gamma$ is increasing in $x >0$, we have:
\begin{equation}
\label{eq23}
\frac{\big\| \eJ^{-1/2}_m \sum^{m+ {k}^0_m}_{i=m+1} g_\tau (Y_i - \eX_i^\top \widehat{\eb}_m) \eX_i \big\|_\infty }{ z(m, \widetilde{k}_m, \gamma)} \leq C < \infty ,
\end{equation}
with probability converging to  1 as $m \rightarrow \infty $.\\
For the random vector given by (\ref{eq22}) it remains to study  $ \sum^{m+\widetilde{k}_m }_{i=m+{k}^0_m+1}  g_\tau (Y_i - \eX_i^\top \widehat{\eb}_m)  \eX_i$. For this, taking into account the convergence rate of $\widehat{\eb}_m$ towards $\eb^0$, let us consider the following sum, for $\eu \in \R^p$, with $\|\eu \|_2 <C$: 
\begin{equation}
\label{eq24}
\sum^{m+\widetilde{k}_m }_{i=m+{k}^0_m+1} \eR_i(\eu)=\sum^{m+\widetilde{k}_m }_{i=m+{k}^0_m+1} \big[g_\tau(Y_i - \eX_i^\top(\eb^0+m^{-1/2} \eu))-g_\tau(\varepsilon_i)\big] \eX_i .
\end{equation}
For the mean of the random process $\sum^{m+\widetilde{k}_m }_{i=m+{k}^0_m+1} \eR_i(\eu)$, using Lemma 2 of \cite{Gu-Zou.16} and relation (\ref{eq24}), we have that:
\[
\eE \big[\sum^{m+\widetilde{k}_m }_{i=m+{k}^0_m+1} \eR_i(\eu) \big] = \sum^{m+\widetilde{k}_m }_{i=m+{k}^0_m+1} C_i \eX_i \eX_i^\top \big(\eb^1 - \eb^0 +m^{-1/2} \eu \big) ,
\]
with $C_i$ a constant such that $  |C_i| \in [2\underline{c}, 2 \bar{c}]$. \\
Using the supposition that:    $m^{-s} \big\| \sum^{m+k^0+m^s }_{i=m+k^0+1} C_i \eX_i \eX_i^\top  \big\|_\infty > C >0$ and since $\widetilde{k}_m= k^0_m +m^s$, we get:  
\begin{equation}
\label{eq25}
m^{-s} \big\| \sum^{m+\widetilde{k}_m }_{i=m+{k}^0_m+1} C_i \eX_i \eX_i^\top  \big\|_\infty > C >0.
\end{equation}
If $\| \eb^1 -\eb^0\|_2 > C > 0$, then, relation (\ref{eq25}) implies that
\[
\frac{1}{\widetilde{k}_m - {k}^0_m} \bigg\| \sum^{m+\widetilde{k}_m }_{i=m+{k}^0_m+1} \eE[\eR_i(\eu)] \bigg\|_\infty > C > 0,
\]
and using the LLN we obtain:
\begin{equation}
\label{eq26}
\bigg\| \sum^{m+\widetilde{k}_m }_{i=m+{k}^0_m+1} \eR_i(\eu) \bigg\|_\infty = O_\PP \big(\widetilde{k}_m - {k}^0_m \big)=O_\PP(m^s).
\end{equation}
If $ \eb^1 -\eb^0 {\underset{m \rightarrow \infty}{\longrightarrow}} \textbf{0}_p$ such that $m^{1/2} \| \eb^1 -\eb^0\|_2 {\underset{m \rightarrow \infty}{\longrightarrow}} \infty  $ then,
\begin{equation}
\label{eq27}
\bigg\| \sum^{m+\widetilde{k}_m }_{i=m+{k}^0_m+1} \eR_i(\eu) \bigg\|_\infty = O_\PP \big((\widetilde{k}_m - {k}^0_m) \| \eb^1 -\eb^0\|_2 \big).
\end{equation}
On the other hand, by the Central Limit Theorem, we have:
\begin{equation}
\label{eq28}
\sum^{m+\widetilde{k}_m }_{i=m+{k}^0_m+1}  g_\tau(\varepsilon_i) \eX_i= O_\PP \big(\widetilde{k}_m - {k}^0_m \big)^{1/2}=O_\PP(m^{s/2}).
\end{equation}
Therefore, combining (\ref{eq24}), (\ref{eq26}), (\ref{eq27}), (\ref{eq28}), we obtain:
\begin{equation}
\begin{split}
\frac{\big\| \eJ^{-1/2}_m\sum^{m+\widetilde{k}_m }_{i=m+{k}^0_m+1}  g_\tau (Y_i - \eX_i^\top (\eb^0+m^{-1/2} \eu)) \eX_i\big\|_\infty }{ z(m, \widetilde{k}_m, \gamma)}  \qquad \qquad\\  
=\frac{O_\PP(m^{s/2})+O_\PP(m^{s-1/2} \|m^{1/2}(\eb^1-\eb^0) \|_\infty)}{ z(m, \widetilde{k}_m, \gamma)} {\underset{m \rightarrow \infty}{\longrightarrow}} \infty
\end{split}
\label{ter}
\end{equation}
(see also relation (4.25) of \cite{Ciuperca-17}). Because the convergence rate of $\widehat{\eb}_m$ to $\eb^0$ is of order $m^{-1/2}$, relation (\ref{ter}) implies:
\begin{equation}
\label{terb}
\frac{\big\| \eJ^{-1/2}_m\sum^{m+\widetilde{k}_m }_{i=m+{k}^0_m+1}  g_\tau (Y_i - \eX_i^\top \widehat{\eb}_m) \eX_i\big\|_\infty }{ z(m, \widetilde{k}_m, \gamma)} \overset{\PP} {\underset{m \rightarrow \infty}{\longrightarrow}} \infty .
\end{equation}
The theorem follows by combining relations (\ref{eq23}) and (\ref{terb}). 
\hspace*{\fill}$\blacksquare$  \\

{\bf Proof of Proposition  \ref{Propositon 1A}.} 
The Karush-Kuhn-Tucker optimality conditions for the adaptive LASSO expectile estimator $\widehat{\eb}^*_m$ are:
\begin{equation}
\label{KKT_i}
(i) \quad \forall j \in \widehat{\cal A}^{*}_m, \quad \textrm{we have: } \sum^m_{i=1} g_\tau (Y_i - \eX_i^\top \widehat{\eb}^*_m)X_{ij}=m \lambda_m \widehat{\omega}_{m,j} \textrm{sgn}(\widehat{\beta}^*_{m,j}),
\end{equation}
\begin{equation}
\label{KKT_ii}
(ii) \quad \forall j \in \widehat{\cal A}^{*^c}_m, \quad \textrm{we have: } \bigg| \sum^m_{i=1} g_\tau (Y_i - \eX_i^\top \widehat{\eb}^*_m)X_{ij} \bigg| \leq m \lambda_m \widehat{\omega}_{m,j},
\end{equation}
with $sgn(x)$ the sign of $x$: $sgn(x)=x/|x|$ for $x \neq 0$ and $sgn(0)=0$.\\
Taking into account the convergence rate of the expectile estimator $\widehat{\eb}_{m,{\cal A}^0}$ towards $\eb^0_{{\cal A}^0}$ and the supposition $m^{1/2} \lambda_m=o(1)$ of assumption (A5), we have for all $j \in {\cal A}^0$:
\begin{equation}
\label{tr}
m \lambda_m \widehat{\omega}_{m,j} =O_\PP(m \lambda_m)=o_\PP(m^{1/2}).
\end{equation}
From relation (\ref{KKT_i}) we get from all $j \in \widehat{\cal A}^{*}_m \cap {\cal A}^0$:
\begin{equation}
\label{eq_8j}
 \sum^m_{i=1} g_\tau \big(\varepsilon_i - \eX^\top_{i,{\cal A}^0}( \widehat{\eb}^*_m -\eb^0)_{{\cal A}^0}\big)X_{ij}=m \lambda_m \widehat{\omega}_{m,j} \textrm{sgn}(\widehat{\beta}^*_{m,j}).
\end{equation}
Using relation (\ref{eq3}) and the convergence in probability of $\widehat{\eb}^*_{m, {\cal A}^0}$ to $\eb^0_{{\cal A}^0}$, we have:
\[
g_\tau \big(\varepsilon_i - \eX^\top_{i,{\cal A}^0}( \widehat{\eb}^*_m -\eb^0)_{{\cal A}^0}\big)=g_\tau(\varepsilon_i)-h_\tau(\varepsilon_i)\eX^\top_{i,{\cal A}^0} (\widehat{\eb}_m - \eb^0)_{{\cal A}^0}+o_\PP \big( \eX^\top_{i,{\cal A}^0} (\widehat{\eb}_m - \eb^0)_{{\cal A}^0}\big).
\]
Taking into account this last relation in (\ref{eq_8j}), we obtain for all $j \in \widehat{\cal A}^{*}_m \cap {\cal A}^0$:
\begin{eqnarray}
\label{eq_10j}
m \lambda_m \widehat{\omega}_{m,j} \textrm{sgn}(\widehat{\beta}^*_{m,j})= \sum^m_{i=1} g_\tau(\varepsilon_i)- \sum^m_{i=1} h_\tau(\varepsilon_i)\eX^\top_{i,{\cal A}^0} (\widehat{\eb}_m - \eb^0)_{{\cal A}^0} +\sum^m_{i=1}o_\PP \big( \eX^\top_{i,{\cal A}^0} (\widehat{\eb}_m - \eb^0)_{{\cal A}^0}\big). \qquad
\end{eqnarray}
Taking into account (\ref{KKT_ii}), (\ref{tr}) and (\ref{eq_10j}) we obtain by the same arguments as in the proof of Proposition \ref{Propositon 1} that:
\begin{equation*}
\begin{split}
\label{eq7j}
(\widehat{\eb}_m - \eb^0)_{{\cal A}^0} &= \bigg(\frac{1}{m}    \sum^m_{i=1} h_\tau(\varepsilon_i)\eX_{i,{\cal A}^0}\eX^\top_{i,{\cal A}^0}\bigg)^{-1}\bigg( \frac{1}{m} \sum^m_{i=1}g_\tau(\varepsilon_i)\eX_{i,{\cal A}^0} + o_\PP(m^{-1/2})\bigg)\\
& =\mu^{-1}_h  \eO^{-1}_{{\cal A}^0} \frac{1}{m}  \sum^m_{i=1}g_\tau(\varepsilon_i)\eX_{i,{\cal A}^0} +o_\PP(m^{-1/2}),
\end{split}
\end{equation*}
i.e. the claim of the proposition.
\hspace*{\fill}$\blacksquare$  \\

 {\bf Proof of Theorem \ref{Theorem 1(Metrika)}.} 
Taking inspiration from the approach used for the test statistic built on the expectile residuals, let us consider 
 \begin{align*}
 \eu & =\big( {\eu^*_m}^{\top}, {\textbf{0}_{{\cal A}^0 \cap \widehat{\cal A}^{*^c}_m}}^{\top}\big)^\top, \qquad \eu \in \R^{|{\cal A}^0|}, \quad \|\eu \|_2 \leq C,\\
 \rr^0_{m,k}(\eu)& \equiv  \sum^{m+k}_{i=m+1}  \bigg( g_\tau( Y_i - \eX_{i,{\cal A}^0}^\top(\eb^0_{{\cal A}^0} +m^{-1/2}\eu)) - g_\tau(  Y_i - \eX_{i,{\cal A}^0}^\top\eb^0_{{\cal A}^0}) \bigg) \eX_{i,{\cal A}^0}.
 \end{align*}
 
\noindent \underline{If $k$ small}, using Lemma 2 of \cite{Gu-Zou.16} and since $\gamma <1/2$, we have:
\[
\frac{\|\rr^0_{m,k}((\eu)) \|_2}{z(m,k,\gamma)} \leq \frac{2^\gamma k m^{-1/2} \max_{m+1 \leqslant i \leqslant\leqslant m+k} \| \eX_i\|_2}{m^{1/2}(k/m)^\gamma}=o(1).
\] 
\underline{If $k$ large}. By Lemma \ref{Lemma 4.2 (SQA)} we have:
\begin{equation}
\label{eq26bis}
\rr^0_{m,k}(\eu)=\eE [ \rr^0_{m,k}(\eu)]+O_\PP(R_{m,k}),
\end{equation}
with $R_{m,k}= \big( k m^{-1}\big)^{1/2}  (\log k)^{1/2}$. For  the expectation in the   right-hand side of (\ref{eq26bis}) we have, combining relation (\ref{eq3}), together with the supposition that $\eE[g_\tau(\varepsilon_i)]=0$ of (A3), with assumption (A1), that:
\begin{equation}
\begin{split}
\label{eq27bis}
\eE[\rr^0_{m,k}(\eu)]&= - \mu_h \sum^{m+k}_{i=m+1} \eX_{i,{\cal A}^0} \eX_{i,{\cal A}^0}^\top m^{-1/2} \eu(1+o(1))\\
&=- m^{-1/2} \mu_h \sum^{m+k}_{i=m+1} \eX_{i,{\cal A}^0} \eX_{i,{\cal A}^0}^\top \eu +o(k m^{-1/2}).
\end{split}
\end{equation}
Since $\widehat{\cal A}^*_m \subseteq {\cal A}^0$ and $\|\widehat{\eb}^*_m- \eb^0 \|_2=O_\PP(m^{-1/2})$, then we can take $\eu=m^{1/2}(\widehat{\eb}^*_{m,{\cal A}^0} - \eb^0_{{\cal A}^0})$ in the definition of $\rr^0_{m,k}$. Thus,  we obtain:
\begin{equation}
\label{eq31}
\rr^0_{m,k}\big(m^{1/2}(\widehat{\eb}^*_{m,{\cal A}^0} - \eb^0_{{\cal A}^0})\big)=\sum^{m+k}_{i=m+1}  \bigg(g_\tau(\varepsilon_i- \eX_i^\top(\widehat{\eb}^*_m - \eb^0))-g_\tau(\varepsilon_i) \bigg)\eX_{i,{\cal A}^0} .
\end{equation}
On the other hand, combining relations (\ref{eq26bis}) and (\ref{eq27bis}) we have:
\begin{align}
\rr^0_{m,k}\big(m^{1/2}(\widehat{\eb}^*_{m,{\cal A}^0} - \eb^0_{{\cal A}^0})\big)& = - \mu_h \sum^{m+k}_{i=m+1} \eX_{i,{\cal A}^0}  \eX_{i,{\cal A}^0}^\top  (\widehat{\eb}^*_{m,{\cal A}^0} - \eb^0_{{\cal A}^0})+O_\PP(R_{m,k})+o_\PP(k m^{-1/2}) \nonumber \\
& = - k \mu_h \eO_{m, {\cal A}^0} (\widehat{\eb}^*_m - \eb^0)_{{\cal A}^0}+O_\PP(R_{m,k})+o_\PP(k m^{-1/2}).   \label{eq32}
\end{align}
By Proposition \ref{Propositon 1A}  we have that
\[
\widehat{\eb}^*_{m,{\cal A}^0}=\eb^0_{{\cal A}^0} +\mu_h^{-1}\frac{\eO^{-1}_{{\cal A}^0}}{m} \sum^m_{i=1} g_\tau(\varepsilon_i) \big( \eX_{i,{\cal A}^0} +o_\PP(m^{-1/2})\big).
\]
Then,  relation (\ref{eq32}) becomes:
\[
- \frac{k}{m} \eO_{m, {\cal A}^0} \eO^{-1}_{{\cal A}^0}\sum^m_{i=1} g_\tau(\varepsilon_i)  \eX_{i,{\cal A}^0}+O_\PP(R_{m,k})+o_\PP(k m^{-1/2}).
\]
Thus, taking also into account relation (\ref{eq31}) combined with  $\sum^m_{i=1} g_\tau(\varepsilon_i) \eX_{i,{\cal A}^0}=O_\PP(m^{1/2})$ given by relation (\ref{eq6}), with  assumption (A2), we obtain:
\[
\eJ^{-1/2}_{m, {\cal A}^0} \sum^{m+k}_{i=m+1} [g_\tau(\widehat{\varepsilon}^*_i)-g_\tau(\varepsilon_i)]\eX_{i,{\cal A}^0}=-  \frac{k}{m}  \eJ^{-1/2}_{m, {\cal A}^0}\sum^m_{i=1} g_\tau(\varepsilon_i)\eX_{i,{\cal A}^0}+O_\PP(R_{m,k})+o_\PP(k m^{-1/2}),
\] 
from where
\begin{equation}
\begin{split}
\label{eq33}
\eJ^{-1/2}_{m,{\cal A}^0}  \sum^{m+k}_{i=m+1}  g_\tau(\widehat{\varepsilon}^*_i) \eX_{i,{\cal A}^0}= \eJ^{-1/2}_{m, {\cal A}^0} \sum^{m+k}_{i=m+1} g_\tau(\varepsilon_i)\eX_{i,{\cal A}^0}- \frac{k}{m}\eJ^{-1/2}_{m, {\cal A}^0}\sum^m_{i=1} g_\tau(\varepsilon_i)\eX_{i,{\cal A}^0}\\
+O_\PP(R_{m,k})+o_\PP(k m^{-1/2}).
\end{split}
\end{equation}
On the other hand, for the two remainders of relation (\ref{eq33}), as $m \rightarrow \infty$, we have: 
\begin{description}
	\item[] $\displaystyle{\frac{o(k m^{-1/2})}{z(m,k,\gamma)}   \longrightarrow  0}$, taking into account relation (23) of \cite{Ciuperca-18},
	\item $\displaystyle{\frac{R_{m,k}}{z(m,k,\gamma)}   \longrightarrow 0}$, taking into account relation (\ref{eq20}).
\end{description}
The end of the proof is similar to that of Theorem  \ref{Theorem 2.1(SQA)}.
\hspace*{\fill}$\blacksquare$  \\

{\bf Proof of Corollary \ref{Corollary 1(Metrika)}. } 
The proof is similar to that of Corollary 1 in  \cite{Ciuperca-18} and it is omitted.
\hspace*{\fill}$\blacksquare$  \\

{\bf Proof of Theorem  \ref{Theorem 2(Metrika)} } 
In order to expose the major points of the argument, the proof is presented for   $s=1$ only, the other cases being similar.  Consider the observation $\widetilde{k}_m=k^0_m+m$. For the denominator of the statistic $\Theta(m,\widetilde{k}_m,\gamma)$ we make the decomposition: 
\[
\eJ^{-1/2}_{m,{\cal A}^0}  \sum^{m+\widetilde{k}_m}_{i=m+1} g_\tau(\widehat{\varepsilon}^*_i) \eX_{i,{\cal A}^0} 
= \eJ^{-1/2}_{m, {\cal A}^0} \bigg( \sum^{m+k^0_m}_{i=m+1}g_\tau(\widehat{\varepsilon}^*_i) \eX_{i,{\cal A}^0}+  \sum_{i=m+k^0_m+1}^{m+\widetilde{k}_m}g_\tau(\widehat{\varepsilon}^*_i) \eX_{i,{\cal A}^0} \bigg).
\]
By the proof of  Theorem \ref{Theorem 1(Metrika)} we have with the probability converging to  1 as $m \rightarrow \infty $, that:
\[
\bigg\|\eJ^{-1/2}_{m, {\cal A}^0}  \sum^{m+k^0_m}_{i=m+1}g_\tau(\widehat{\varepsilon}^*_i) \eX_{i,{\cal A}^0} \bigg\|_\infty /z(m,\widetilde{k}_m,\gamma) \leq C.
\]
Taking into account the convergence rate of $\widehat{\eb}^*_m$ towards $\eb^0$ (see Theorem 2.1 of \cite{Ciuperca.19}) and the sparsity property of $\widehat{\eb}^*_m$, let  the random process:
\[
\textbf{R}^0_i(\eu^*_m) \equiv  \bigg(g_\tau \big( Y_i -\eX^\top_{i, \widehat{{\cal A}}^*_m}(\eb^0_{\widehat{\cal A}^*_m}+m^{-1/2} \eu^*_m) \big) - g_\tau(\varepsilon_i)  \bigg)\eX_{i,{\cal A}^0},
\]
with $\eu^*_m \in \R^{|\widehat{{\cal A}}^*_m|}$, $\|\eu^*_m \|_2 \leq C$. Taking into account relation (\ref{eq25bis}), we have:
\[
\sum^{m+\widetilde{k}_m}_{i=m+k^0_m+1} \textbf{R}^0_i(\eu^*_m) =\sum^{m+\widetilde{k}_m}_{i=m+k^0_m+1}  \bigg(g_\tau \big( Y_i -\eX^\top_{i,  {{\cal A}^0}}(\eb^0_{{\cal A}^0}+m^{-1/2} \eu) \big) - g_\tau(\varepsilon_i)  \bigg)\eX_{i,{\cal A}^0}
\]
\[
- \sum^{m+\widetilde{k}_m}_{i=m+k^0_m+1}  \big[g_\tau \big( Y_i -\eX^\top_{i,  {{\cal A}^0}}(\eb^0_{{\cal A}^0}+m^{-1/2} \eu) \big) - g_\tau\big( Y_i -\eX^\top_{i, {{\cal A}^0}}(\eb^0_{{\cal A}^0}+m^{-1/2} \eu) -\eX^\top_{i,{\cal A}^0 \cap \widehat{\cal A}^{*^c}_m}\eb^0_{{\cal A}^0 \cap \widehat{\cal A}^{*^c}_m} \big)   \big]\eX_{i,{\cal A}^0}
\]
\[
\equiv D_1(\eu) - D_2(\eu), \qquad \qquad \qquad \qquad
\]
with $\eu=(\eu^*_m,\textbf{0}_{|{\cal A}^0 \cap \widehat{\cal A}^{*^c}_m|})$, $\eu \in \R^{|{\cal A}^0|}$.\\
We start with the study of $D_2(\eu)$. 
For $i=m+k^0_m+1, \cdots m+ \widetilde{k}_m$, let be the probability $p_{m,i} \equiv \PP \bigg[g_\tau \big( Y_i -\eX^\top_{i,  {{\cal A}^0}}(\eb^0_{{\cal A}^0}+m^{-1/2} \eu) \big) = g_\tau\big( Y_i -\eX^\top_{i, {{\cal A}^0}}(\eb^0_{{\cal A}^0}+m^{-1/2} \eu) -\eX^\top_{i,{\cal A}^0 \cap \widehat{\cal A}^{*^c}_m}\eb^0_{{\cal A}^0 \cap \widehat{\cal A}^{*^c}_m} \big)   \bigg]$. Since $\lim_{m \rightarrow \infty} \PP[{\cal A}^0=\widehat{\cal A}^*_m]=1$, together with the fact that  $g_\tau(.)$ is a continuous Borel function, we get that: $\lim_{m \rightarrow \infty} p_{m,i} =1$ for any $i=m+k^0_m+1, \cdots m+ \widetilde{k}_m$.\\
For $i=m+k^0_m+1, \cdots m+ \widetilde{k}_m$, let be the independent random variables $G_i \equiv g_\tau \big( Y_i -\eX^\top_{i,  {{\cal A}^0}}(\eb^0_{{\cal A}^0}+m^{-1/2} \eu) \big) - g_\tau\big( Y_i -\eX^\top_{i,  {{\cal A}^0}}(\eb^0_{{\cal A}^0}+m^{-1/2} \eu) -\eX^\top_{i,{\cal A}^0 \cap \widehat{\cal A}^{*^c}_m}\eb^0_{{\cal A}^0 \cap \widehat{\cal A}^{*^c}_m} \big)  $ which takes the value 0 with probability $p_{m,i}$. 
The variance of  $G_i$ exists and  is bounded.\\
Then, by the  Bienaymé-Tchebychev inequality, taking into account also assumption  (A1), we obtain that for  all $ j \in {\cal A}^0$:
\[
\frac{1}{\widetilde{k}_m - k^0_m} \sum^{m+\widetilde{k}_m}_{i=m+k^0_m+1} X_{ij} G_i =\frac{1}{m} \sum^{2m+k^0_m}_{i=m+k^0_m+1} X_{ij} G_i  \overset{\PP} {\underset{m \rightarrow \infty}{\longrightarrow}} 0.
\]
Thus,
\[
\big\|\sum^{m+\widetilde{k}_m}_{i=m+k^0_m+1} \eX_{i, {\cal A}^0} G_i  \big\|_\infty=o_\PP(\widetilde{k}_m - k^0_m)=o_\PP(m),
\]
which implies $\|  D_2(\eu)\|_\infty =o_\PP(m) $.\\
We now study $ D_1(\eu)$. As in the proof of Theorem \ref{Theorem 2.2(SQA)} we have:
\[
\eE\bigg[\bigg\|\sum^{m+\widetilde{k}_m}_{i=m+k^0_m+1}  \big[g_\tau \big( Y_i -\eX^\top_{i,  {{\cal A}^0}}(\eb^0_{{\cal A}^0}+m^{-1/2} \eu) \big) - g_\tau(\varepsilon_i)   \big] \eX_{i, {\cal A}^0} \bigg\|_\infty  \bigg]\]
\[
=O_\PP \bigg(\bigg\|\sum^{m+\widetilde{k}_m}_{i=m+k^0_m+1}C_i \eX_{i, {\cal A}^0} \eX_{i,A}^t (\widetilde{\eb}^1 -\widetilde{\eb}^0 +m^{ -1/2} \widetilde{\eu} ) \bigg\|_\infty  \bigg)
\]
and the end of the proof is similar to that of Theorem  \ref{Theorem 2.2(SQA)}.
\hspace*{\fill}$\blacksquare$  \\

{\bf Proof of Corollary \ref{Corollaire 2(Metrika)}. } 
By the sparsity property of $\widehat{\eb}^*_m$ we have: $\eJ^{-1/2}_{m, \widehat{\cal A}^*_m} \sum^{m+\widetilde{k}_m}_{i=m+1}  g_\tau(Y_i - \eX_i^\top \widehat{\eb}^*_m )\eX_{i, \widehat{\cal A}^*_m} (1+o_\PP(1))= \eJ^{-1/2}_{m,{\cal A}^0}  \sum^{m+\widetilde{k}_m}_{i=m+1} g_\tau(\widehat{\varepsilon}^*_i) \eX_{i,{\cal A}^0}(1+o_\PP(1))$ and the Corollary results  taking into account  Theorem \ref{Theorem 2.2(SQA)}. 
\hspace*{\fill}$\blacksquare$  \\


\end{document}